\documentclass[journal]{IEEEtran}
\usepackage{dirtytalk}
\usepackage{booktabs}
\usepackage{adjustbox}
\usepackage{multirow, makecell}
\usepackage{multicol}
\usepackage{subfig}
\usepackage{amsmath,amssymb,amsfonts, amsthm}
\usepackage{xcolor,colortbl}
\usepackage{algorithm}
\usepackage{hyperref}
\usepackage{xnewcommand}
\usepackage{caption}
\usepackage[noend]{algpseudocode}
\usepackage{tabularx}
\usepackage{graphics}
\usepackage{scalerel}
\usepackage{setspace}
\usepackage{pifont}% http://ctan.org/pkg/pifont
\newcommand{\cmark}{\ding{51}}

\newcommand{\proc}[1]{\ifmmode\mbox{\textsc{#1}}\else\textsc{#1}\fi}
\newtheorem*{remark}{Remark}
\definecolor{maroon}{cmyk}{0,0.87,0.68,0.32}
\newcommand{\eg}{\mbox{{\em e.g.}}}

\newcommand{\ie}{\mbox{{\em i.e.}}}
\newcommand{\viz}{\mbox{{\em viz.}}}

\begin{document}

\title{Resilient Cooperative Adaptive Cruise Control for Autonomous Vehicles Using Machine Learning}
%Real-time Resiliency for Cooperative Adaptive Cruise Control Using Machine Learning}

\author{Srivalli Boddupalli,~\IEEEmembership{Student Member,~IEEE},
        Akash Someshwar Rao,
        %Sattanaathan Thayumanan,
        and~Sandip~Ray,~\IEEEmembership{Senior Member,~IEEE}% <-this % stops a space
\thanks{The authors are affiliated to the Department
of Electrical and Computer Engineering at the University of Florida, Gainesville, FL 32611, USA. e-mail: bodsrivalli12@ufl.edu, akash.someshwarr@ufl.edu, sandip@ece.ufl.edu.}}

\maketitle

\begin{abstract}

Cooperative Adaptive Cruise Control (CACC) is a fundamental connected vehicle application that extends Adaptive Cruise Control  by  exploiting vehicle-to-vehicle (V2V) communication.  CACC is a crucial ingredient for numerous autonomous vehicle functionalities including platooning, distributed route management, etc.  Unfortunately, malicious V2V communications can subvert CACC, leading to string instability and road accidents. In this paper, we develop a novel resiliency infrastructure, RACCON, for detecting and mitigating V2V attacks on CACC. RACCON  uses  machine  learning  to develop an on-board prediction model that captures anomalous vehicular responses and performs mitigation in real time. RACCON-enabled vehicles can exploit the  high efficiency of CACC without compromising safety, even under potentially adversarial scenarios.  We present  extensive experimental evaluation to demonstrate the efficacy of RACCON.

\end{abstract}

\begin{IEEEkeywords}
Connected and autonomous vehicles, V2X communication, anomaly detection, security
\end{IEEEkeywords}

\section{Introduction}
\label{sec:Intro}

Recent years have seen proliferation of  electronics and software in automotive systems targeted towards increasing autonomy.
%Modern automotive systems include hundreds of connected Electronic Control Units (ECUs), a variety of sensors and actuators, $3$-$5$ in-vehicle networks, several miles of cable, and several hundred megabytes of software code.    
Autonomous features hold the promise of dramatically increasing transportation efficiency and road safety by reducing and eventually eliminating human errors \cite{nhtsa}. However, an undesired side-effect is the increased vulnerability of Connected and Autonomous Vehicles (CAVs) to cyber-security threats. %However, an upshot is  increase in their vulnerability to  cyber-attacks.
Recent research has shown that it is possible, even straightforward, to mount cyber-attacks that compromise a vehicle and control its driving functionality \cite{miller2015,checkoway2011,koscher2010}.  Increasing dependence of critical vehicular operations on communication with the external world will exacerbate this situation by creating  larger attack surfaces. This increases the attacker's ability to compromise the vehicle causing catastrophic impact.   Consequently, the proliferation and even adoption of CAVs depends critically on our ability to  mitigate such attacks.

An important feature  of autonomous vehicles is the ability to interact with other vehicles (V2V), the transportation infrastructure (V2I), and  devices connected to the Internet (V2IoT).  Vehicular communications, collectively referred to as V2X, form a key constituent of several  emergent applications including platooning, cooperative route management, intersection management, cooperative collision detection, etc. Unfortunately, V2X also enables a large class of adversarial opportunities: an adversary can easily create disruption by manipulating communicated messages through mutation, misdirection, or jamming.   For example, in platooning, the adversary may cause an accident by simply sending misleading acceleration directive while braking \cite{platoon-security}.

In this paper, we develop an infrastructure for systematically integrating resiliency against communication attacks on V2V applications.  Our focus is a fundamental application of vehicular communications: Cooperative Adaptive Cruise Control (CACC). CACC is an extension of Adaptive Cruise Control (ACC); %in addition to the RADAR/LIDAR measurements used in ACC to derive the relative velocity and headway from the vehicle in front, CACC  also accounts for the (intended) acceleration of the preceding vehicle.  
Adaptive Cruise Control (ACC) uses RADAR/LIDAR measurements to derive relative velocity and headway from the vehicle in front. Additionally, CACC also accounts for the preceding vehicle’s (intended) acceleration.
The acceleration is communicated through V2V messages, typically as Dedicated Short Range Communication (DSRC) \cite{ModelPredictiveCACC}.  CACC is a key component of several  connected car applications such as vehicle platooning, cooperative on-ramp merging, etc.  Attacks on CACC can disrupt traffic movement, cause catastrophic accidents, and bring down the transportation infrastructure.

Our framework, RACCON (for ``Resilient Cooperative Adaptive Cruise Control''), is a real-time anomaly detection and mitigation system for communication attacks on CACC.  The key idea  is to use machine learning (ML) to develop an on-board prediction model for estimating the response of the following vehicle given normal (benign) patterns of V2V input messages. This enables the detection of anomalies in the vehicle's responses resulting from potentially malicious communications. RACCON involves two cooperative components: (1) an on-board architecture installed in vehicles participating in CACC that enables the follower vehicle (also called \textit{ego vehicle}) to perform real-time anomaly detection and mitigation; and (2) an offline cloud-based infrastructure for construction of prediction models.

%{\em To our knowledge, RACCON represents the first resiliency framework on CACC that can uniformly address a comprehensive spectrum of communication attacks.} 
%We provide extensive experimental results to demonstrate the viability of RACCON.

%The paper makes several important  contributions. First, our anomaly detection accounts for computational constraints of automotive electronics, compatibility with the existing CACC architecture, and real-time constraints imposed by connected car applications.  Second, our approach can address an elaborate set of adversaries in the connected car ecosystem, including  man-in-the-middle (MITM) attack, wormhole attack, Sybil, Denial-of-Service (DoS), and many others.  Furthermore, it accounts for the natural differences in communication patterns among a variety of driving scenarios, road conditions, etc.  {\em This is in stark contrast to related control-theory approaches on V2V security \cite{CACCFaultDetection, SecForSafety} that require detailed, continuous models of  vehicular and adversarial functionalities.}  Finally, in addition to detecting anomalies, our approach enables real-time mitigation that enables the ego vehicle to continue to employ CACC even under certain detected adversarial scenarios instead of  degrading to Adaptive Cruise Control (ACC) without cooperation.

The paper makes several important contributions.  First, unlike related approaches that focus on {\em detection} of CACC attacks (see Section \ref{sec:Related}), RACCON represents  the first framework that also enables {\em real-time resiliency}.
%Addressing mitigation requires not only effective detection of anomalous V2X communication but also comprehending the response of the CACC controller to such communication, estimating the impact of the response to safety and efficiency, and computing alternate response to minimize the severity of the impact.  
%%% for detecting V2X adversaries in CACC that accounts for \textit{real-time response} of the controller and performs mitigation.  While there has recently been several related approaches on ML-based detection of CACC anomalies (see Section \ref{sec:Related}), none of these approaches accounts for real-time mitigation.  
%The ability to comprehend real-time response of the vehicle is crucial to developing cyber-resilience for cooperative autonomous vehicle (CAV) applications.  
Second, our framework provides high flexibility through attack-agnostic defense against an elaborate set of adversaries in the connected car ecosystem, including  man-in-the-middle (MITM) attack, wormhole attack, Sybil, Denial-of-Service (DoS), etc.  
%This is in stark contrast to related control-theory approaches on V2X security \cite{CACCFaultDetection, SecForSafety} and existing ML detection techniques \cite{9061150, 9062798, jagielski2018threat}.  
RACCON also accounts for the natural differences in communication patterns among a variety of driving scenarios, road conditions, etc. Finally, our work represents the most comprehensive  experimental evaluation to date on vulnerabilities in CACC, impact of attacks on target vehicles, and the quality of resiliency provided by the security architecture. In addition to showcasing confidence in our approach, we believe the experimental framework will serve as a roadmap for evaluation of resiliency in other CAV applications.

The remainder of the paper is organized as follows.  Section~\ref{sec:Background} provides relevant background in V2X, cooperative vehicular applications, and CACC.  We introduce RACCON in Section \ref{sec:IntroRACCON} and explain the design constraints induced by the requirements for real-time detection and mitigation.  Section \ref{sec:RACCONFunc} presents details of the RACCON architecture and implementation.  A unique contribution of the paper is the extensive evaluation performed to demonstrate the efficacy of RACCON.  Sections \ref{sec:ExpSetup} through \ref{sec:DetectorSubversion} explain our experimental results.  We discuss related work in Section \ref{sec:Related} and conclude in Section \ref{sec:Concl}.

\section{Background}
\label{sec:Background}

\subsection{Connected Vehicle Applications and V2X Communications}

Connected autonomous vehicle (CAV) applications exploit Internet connectivity  to enhance driving efficiency, safety, mobility, and sustainability \cite{automated-highway}.  
%In particular, connected vehicle technology can provide real-time information about the surrounding traffic condition and the traffic management center's decisions. 
With the increasing proliferation and speed of Internet connectivity, several such applications have emerged in the past decade.  
%US DOT has developed a Connected Vehicle Reference Architecture (CVRA) to help guide deployment of components by road operators and automotive, highway, and aftermarket equipment manufacturer and service providers.   
Today, connected vehicle applications constitute some of the core components of R\&D around autonomous driving.  Some applications currently designed  include platooning, dynamic cooperative route management (DCRM), automated collision detection, cooperative automated on-ramp merging, etc.

V2X is an essential centerpiece of all CAV applications. DSRC is a popular communication scheme that enables V2X. It is a modified version of the IEEE 802.11p Wide Local Area Network (WLAN) protocol, designed for vehicular ad-hoc networks comprising high mobility nodes. US Federal Communications Commission (FCC) has allocated a dedicated bandwidth of $75$MHz in the $5.850$-$5.925$GHz band to DSRC. In an effort to expedite the deployment of connected vehicle technologies, United States Department of Transportation put forth a proposal in 2016 mandating integration of DSRC devices on all new light-duty vehicles produced in USA.  

\subsection{ACC and CACC Overview}
\label{subsec:cacc}

\begin{figure}
\begin{center}
\begin{minipage}{\columnwidth}
\includegraphics[width=\columnwidth]{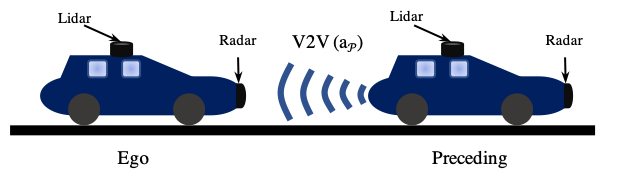}
\end{minipage} 
\end{center}
\caption{{\small CACC Overview. Acceleration provided by V2V. Instantaneous g and $v_{\cal{P}}$ provided by LIDAR or RADAR.}}
\label{fig:cacc}
\end{figure}

Adaptive Cruise Control (ACC) enables a vehicle ${\cal{E}}$ to automatically adjust acceleration and closely follow its preceding vehicle ${\cal{P}}$, while maintaining a safe space gap $g_{\mbox{safe}}$.  Most ACC implementations target a {\em constant time headway}; the goal is to compute $a_{\cal{E}}$ such that ${\cal{E}}$ takes at least time $T_{\mbox{gap}}$ to reach the same position as ${\cal{P}}$, where $T_{\mbox{gap}}$ is a design constant.  The safe space gap $g_{\mbox{safe}}$ is a function of $T_{\mbox{gap}}$, the maximum deceleration capability $D_{\cal{E}}^{\mbox{max}}$ of ${\cal{E}}$, and the velocities $v_{\cal{E}}$ and $v_{\cal{P}}$.  Vehicle ${\cal{E}}$ computes its desired acceleration $a_{\cal{E}}$ using (1)~the inter-vehicle distance $g$ and velocity $v_{\cal{P}}$ of the preceding vehicle ${\cal{P}}$ measured by RADAR/LIDAR; and (2)~the velocity $v_{\cal{E}}$ and acceleration $a_{\cal{E}}$ of ${\cal{E}}$ measured by on-board sensors.   {\em Cooperative Adaptive Cruise Control} (CACC) extends ACC by  using the intended acceleration $a_{\cal{P}}$ of ${\cal{P}}$ in the computation of $a_{\cal{E}}$.  Vehicle ${\cal{P}}$ communicates $a_{\cal{P}}$ through V2V messages (Fig.~\ref{fig:cacc}).  Both ACC and CACC operate in two modes.  If $g > g_{\mbox{safe}}$, they operate in {\em gap control mode}, where ${\cal{E}}$ follows ${\cal{P}}$ as closely as possible; if $g \leq g_{\mbox{safe}}$, they switch to {\em collision avoidance mode} and uses maximum deceleration $D_{\cal{E}}^{\mbox{max}}$.  The use of the preceding vehicle's acceleration enables CACC to maintain a shorter time headway (THW) than ACC, resulting in a more efficient roadway utilization: in a representative  implementation \cite{AMOOZADEH2015110}, CACC uses $T_{\mbox{gap}}$ of $0.55s$ while ACC needs to use $1.2s$.

\subsection{CACC Architecture and A Representative Implementation}

\begin{figure}
\begin{center}
\begin{minipage}{0.85\columnwidth}
\includegraphics[width=\columnwidth]{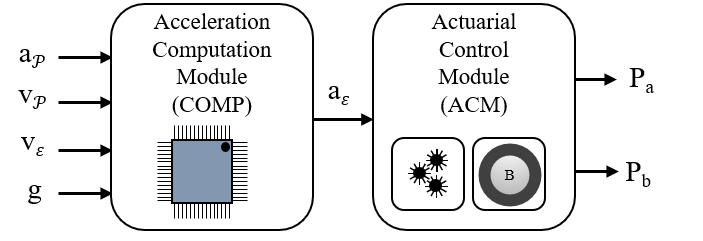}
\end{minipage}
\end{center}
\caption{{\small CACC On-board Architecture.  Acceleration Computation Module (COMP) computes desired acceleration. Actuarial control module (ACM) computes braking pressure and motor torque.}}
\label{fig:cacc-architecture}
\end{figure}

Fig.~\ref{fig:cacc-architecture} shows the key components of an on-board CACC architecture.  While low-level details vary, all implementations to our knowledge constitute two key components: {\em Acceleration Computation Module (COMP)} computes the desired acceleration $a_{\cal{E}}$ of the host vehicle ${\cal{E}}$; {\em Actuarial Control Module (ACM)} manipulates motor output torque or braking pressure to enforce the desired acceleration.  
%The accelerator pedal signal $a$ and braking pressure $P_b$ are determined by a motor torque map and a braking lookup table,  which are obtained from driving test data and included in the on-board hardware.

Although RACCON is oblivious to the underlying CACC implementation, for our evaluation we use representative CACC (and ACC) implementations by Amoozadeh {\em et al.} \cite{AMOOZADEH2015110} shown below.  In the equations,  $a_{\cal{E}}(A)$ and $a_{\cal{E}}(C)$ are the desired accelerations for ACC and CACC respectively, $G_{\mbox{min}}$ is a constant defining a lower bound on space gap, $T_{\mbox{gap}}^A$ and $T_{\mbox{max}}^C$ are constant time headway for ACC and CACC respectively, and $K_a$, $K_v$, and $K_g$ are acceleration, velocity, and position constants.  Amoozadeh {\em et al.} specify the values $K_a= 0.66$, $K_v = 0.99s^{-1}$, $K_g=4.08s^{-2}$, $G_{\mbox{min}}=1m$, $T_{\mbox{gap}}^A=1.2s$, and $T_{\mbox{gap}}^C= 0.55s$.  

\begin{small}
\begin{math}
\begin{aligned}
g_{\mbox{safe}} & = 0.1v_{\cal{E}} + \frac{v_{\cal{E}}^2}{2D_{\cal{E}}^{\mbox{max}}} -\frac{v_{\cal{P}}^2}{2D_{\cal{P}}^{\mbox{max}}} + G_{\mbox{min}} \\
a_{\cal{E}}(A) & = -K_a D_p^{\mbox{max}} + K_v(v_{\cal{P}} - v_{\cal{E}}) + K_g(g - v_{\cal{E}} T_{\mbox{gap}}^A  - G_{\mbox{min}}) \\
a_{\cal{E}}(C) & = K_a a_{\cal{P}} + K_v(v_{\cal{P}} - v_{\cal{E}}) + K_g(g - v_{\cal{E}} T_{\mbox{gap}}^C - G_{\mbox{min}}) 
\end{aligned}
\end{math}
\end{small}

\section{Introduction to RACCON}
\label{sec:IntroRACCON}

\subsection{RACCON as a Service}

At the user level, RACCON is a vehicular service that is enabled with the help of additional on-board hardware (see Section \ref{sec:RACCONFunc}).  We refer to the subscribing vehicle as the {\em ego vehicle},``${\cal{E}}$''; all our evaluations are done from the perspective of an ego vehicle.  When enabled, RACCON collects normal behavior data during ${\cal{E}}$'s operation.  Data from all vehicles with RACCON installed is periodically uploaded to a trusted  cloud server for progressively refining ML models used by the on-board hardware; ${\cal{E}}$ periodically updates the on-board system by downloading the latest ML models.  The communication with cloud is performed when ${\cal{E}}$ is connected to Internet through a trusted network, \eg, when stationary at the owner's residence; on-road connectivity with cloud is not necessary.  During driving operation, if CACC is engaged in ${\cal{E}}$, the on-board hardware automatically detects anomalies in V2V communication from the preceding vehicle, and performs mitigation.  
% Real-time mitigation is a unique feature that distinguishes RACCON  from other ML approaches for anomaly detection in CACC and drives many of our design decisions as described below.

%an ego vehicle engaging in CACC, to provide real-time resiliency while operating in untrusted driving environments. RACCON's on-board detection system captures any adversarial activity resulting in anomalous V2X communication received at the ego vehicle. On detecting anomalous communication, RACCON's mitigation system computes an alternate optimal action for the ego vehicle, neutralizing the potential impact of targeted attacks as well as unintended anomalies. The ML components for detection and mitigation systems are constructed and maintained on a trusted cloud platform. The training data is mined from the data logs stored by the subscribing ego vehicles on road. The ego vehicles periodically download the latest instances of the trained ML models for real-time detection and mitigation.

\subsection{Design Considerations}
\label{sec:DesignConst}

A unique feature that distinguishes RACCON  from related ML approaches for anomaly detection in CACC is \textit{real-time resiliency}.  For our solution to be viable, a number of design constraints must be satisfied. 

%RACCON is an ML-based architecture to provide real-time resiliency of CACC against V2V attacks.  For such a solution to be viable, a number of design constraints must be met.
%Our framework RACCON is a machine learning solution to enable the ego vehicle ${\cal{E}}$ detect and mitigate adversarial communications in real-time. For such a system to be viable, it must satisfy several constraints. 
\begin{itemize}
 \item {\em Basic safety:} ML-based solutions can only provide a ``high probability'' guarantee on prediction accuracy.  Consequently, it is critical that the RACCON mitigation  generates decisions that are {\em safe} (under the assumed threat model), \ie, do not increase the risk of accident in response to a detected anomaly.  
 %Safety should be guaranteed both for (1) a real attack wrongly classified as benign and (2) a benign message classified as an attack.  
\item {\em Flexibility:} The solution should work without modification, for the entire adversarial spectrum. Hence, control-theoretic solutions that require detailed customized  models of adversarial functionalities are infeasible.
\item {\em Limited Computation and Real-time Requirements:} The security system should operate within the computational constraints of an automotive platform and meet real time requirements of CACC application.
\item{\em Small Data Problem and Machine Learning Attacks:} Any ML-based prediction system requires a significant amount of training data.  Obviously, significant attack data  does not exist in real life, a phenomenon we refer to as the {\em small data problem}.  It is critical for the prediction system to be accurate in the presence of limited anomaly data.  Furthermore, the  system must be robust against \textit{detector subversion}, \ie, attacks targeted specifically to fool the prediction system (see Section \ref{sec:DetectorSubversion}).
\end{itemize}

RACCON addresses the resource constraints and real-time requirements by separating the training of ML models from on-road prediction.  A key observation is that the computation-intensive component of machine learning is training predictor models; once a model is created, detection can be performed within the limited resources of automotive ECUs.  Our system includes a cloud-based methodology for training prediction models, while the on-board architecture is responsible for collecting data and performing real-time prediction.  We ensure basic safety by introducing a {\em plausibility checker} which guarantees that RACCON's mitigation cannot compromise safety due to V2V anomalies.  To address the small data problem, we observe that while labelled anomalous/malicious data is limited, data on normal behavior is typically plentiful.  Consequently, we train prediction algorithms to learn {\em normal behavior model} (NBM), \ie, the response of ${\cal{E}}$ to normal (benign) pattern of V2V communications rather than anomalous behavior. The on-board anomaly detector then calculates the degree of deviation from NBM as a measure of the anomaly.  %Finally, to enable resilience against detector subversion attacks, we observe that such attacks require sustained, consistent deviation of  from actual for a continued period of time. 
Finally, for ensuring resiliency under detection subversion attacks, we systematically fine-tune the detection threshold to capture minute anomalies that have a perceptible effect on the safety or efficiency of the target vehicle. As a result, stealthy attacks that indeed subvert the detection system fail to cause any adverse impact on the vehicle.  
% detection any detector subversion attack  to enable resiliency against detector subversion attacks, we tune the anomaly detection threshold to capture any attack that has a perceptible impact on the safety or efficiency of the target vehicle. RACCON ensures that attacks not detected as anomalies have no perceptible effect on the safety or efficiency of the application. 

\subsection{Threat Model}
\label{subsec:ThreatModel}

Given our focus on V2V, our threat model assumes that the attacker can tamper with arbitrary V2V messages.  This includes mutation, denial of delivery, masquerading as a different vehicular or infrastructure entity, message fabrication, etc.  Our framework is oblivious to the source of the attack: it can be a rogue preceding vehicle, a compromised ego vehicle infrastructure component, or an intermediate networking component, \eg, denial of message delivery is possible by compromising the software/hardware component of the ego vehicle or interfering with the communication protocol.  We assume that the RACCON on-board system in the ego vehicle, as well as the actuarial components it controls, are not compromised.  We also assume that the sensory inputs to the ego vehicle are not corrupted.\footnote{There has been significant research showing how vehicular sensors can be compromised \cite{petit2015remote}, \cite{9061150}, \cite{SecForSafety}, \cite{CACCFaultDetection}.  Nevertheless, since the modalities of compromising sensors and V2V are different, it is reasonable in the context of detecting V2V anomalies to assume that the sensory inputs are trusted.}  %Finally,  our infrastructure {\em does not require} real-time communication between the ego vehicle and cloud since transfer of trained models and on-road V2V data can be performed when the vehicle is connected to secure communication channels.

%We consider a threat model under which an adversary can compromise a target ego vehicle engaged in CACC, by exploiting the vulnerabilities in the V2X communication. The on-board sensors and the in-vehicle computing systems are assumed to be trusted. The origin of the attack could be a malicious or compromised preceding vehicle reporting falsified acceleration values. We also consider attacks originating from a Man-In-The-Middle (MITM) adversary intercepting, corrupting or jamming the V2X communication in transit. We note that it is reasonable to assume that an adversary may not be capable of simultaneously compromising communication, sensor systems and in-vehicle electronic control units (ECUs). The security vulnerabilities associated with these systems vary widely from each other. Consequently, the adversarial capabilities and attack mechanisms for exploiting them differ significantly. Defending against an all-powerful adversary may not be viable. Therefore, we confine the scope of our threat model to adversaries compromising V2X communication alone.

\section{RACCON Implementation}
\label{sec:RACCONFunc}

\begin{figure}
\centering
\includegraphics[width=\columnwidth]{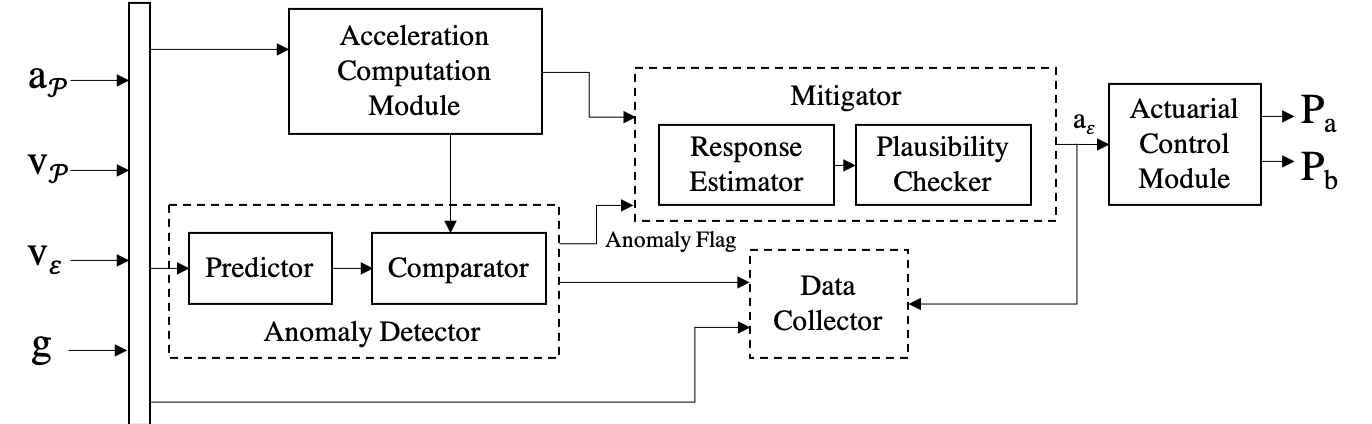}
\caption{{\small RACCON Architecture.  Blocks with dotted line boundaries are components introduced for resiliency.}}
\label{fig:raccon}
\end{figure}

%RACCON includes an on-board architecture for real-time resiliency based on machine learning, supported by a cloud-based infrastructure to build and maintain the relevant ML models.  
Fig.~\ref{fig:raccon} shows the RACCON on-board architecture, and Algorithm~\ref{alg:Robust-CACC} provides a top-level description.  A key insight is that since on-board architecture of most CACC implementations follows the ``template'' from Fig.~\ref{fig:cacc-architecture}, it is possible to develop a streamlined resiliency  architecture by systematically augmenting the template with resiliency components.  RACCON adds three such components:  (1) Anomaly Detector; (2) Mitigator; and (3) Data Collector. 

\begin{algorithm}
\caption{RACCON}
\label{alg:Robust-CACC}
\begin{algorithmic}[1]
\Procedure{RACCON}{$a_{\cal{P}}^{\scaleto{V2V}{2.5pt}},v_{\cal{P}}, v_{\cal{E}}, gap$}
    \setstretch{1.35}
    \State $a_{\cal{P}} \gets a_{\cal{P}}^{\scaleto{V2V}{2.5pt}} $
    \If{V2V communication is lost}
        \State $no\_comm \gets TRUE$
    \EndIf
    \State $a_{\cal{E}}^{\scaleto{pred}{3pt}}  \gets
    \textit{Predictor}() \textbf{\textit{  predictor invoked}}$ 
    \State $a_{\cal{E}}^{\scaleto{C}{3.0pt}} \gets \textit{AccelComp}(a_{\cal{P}},v_{\cal{P}}, v_{\cal{E}}, gap)$
    \State $anmly\_flag \gets \textit{ Comparator}(a_{\cal{E}}^{\scaleto{C}{3.0pt}},a_{\cal{E}}^{\scaleto{pred}{3pt}} )$
    \State $a_{\cal{E}} \gets \textit{Mitigator(anmly\_flag, no\_comm)}$
    \State $throttle, braking \gets \textit {ActuarialControl}(a_{\cal{E}})$
    \State \textit{DataCollector()}
    \State \textbf{return} $throttle, braking$
\EndProcedure
\end{algorithmic}
\end{algorithm}

\subsection{Anomaly Detector}

Anomaly detector checks at each instant $t$ whether the response $a_{\cal{E}}(t)$ of the  \proc{COMP} module of CACC deviates from the expected normal behavior; any such deviation is captured as an anomaly to be passed on to Mitigator.  The detection  comprises the following two modules.
\begin{enumerate}
\item {\bf Predictor} is a machine learning model that is trained offline.  It estimates {\em predicted acceleration value} $a_{\cal{E}}^{pred}(t)$ in real time, taking the same input parameters as \proc{COMP}. Predictor can capture contextual/conditional anomalies, in addition to point anomalies. %However, while the computation $a_{\cal{E}}(t)$ by \proc{COMP} at time $t$ using sensory and V2V inputs at time $t$, the computation of  $a_{\cal{E}}^{pred}(t)$  at time $t$ uses inputs at time $(t-1)$; this enables the system to capture contextual/conditional anomalies, in addition to point anomalies. 

%\footnote{The choice of a single hidden layer feed-forward neural network, is driven by the computational and memory resource constraints of automotive ECUs, the requirements of real-time inference and memory needs of the application, \eg, a more sophisticated model such as Long Short Term Memory incurs significant computation and memory cost, and Recurrent Neural networks may not be required since the CACC controllers are typically memoryless designs that compute acceleration values based on the current inputs.} 

\item {\bf Comparator}  computes the deviation between the predicted value $a_{\cal{E}}^{pred}(t)$ and  $a_{\cal{E}}(t)$; if the deviation is beyond a pre-defined threshold, it is detected as an anomaly.  The detection threshold is  a function of driving conditions and typical velocities of vehicles in a driving environment (See Section \ref{sec:AnomalyThresh}).
\end{enumerate}

%\noindent If no anomaly is detected, $a_{\cal{E}}$ is applied to \proc{ACM} just like regular CACC.  In case of a detected anomaly or interrupted communication, Mitigator is invoked to appropriately handle the anomaly.

\begin{algorithm}
%\color{red}
\caption{Mitigation}
\label{alg:Mitigation}
\begin{algorithmic}[1]

\Procedure{Mitigator}{$anmly\_flag, no\_comm$}
    \setstretch{1.1}
    \If{(anmly\_flag and no\_comm are FALSE)}
        \State \textbf{\textit{operate in normal mode}}
        \State $ a_{\cal{E}} \gets a_{\cal{E}}^{\scaleto{C}{3.0pt}}$
        
    \Else{}
        \State \textbf{\textit{mitigation mode}}
        \State $sensor\_sampling\_frequency \gets F_{max} $
        \State $v_{\cal{P}},gap \gets v_{\cal{P}}^{\scaleto{Fmax}{2.8pt}}, gap^{\scaleto{Fmax}{2.8pt}}$
        \State $a_{\cal{P}} \gets (v_{\cal{P}}(t) - v_{\cal{P}}(t-1)) / \delta T$
        \State $a_{\cal{E}}^{\scaleto{C}{3.0pt}} \gets \textit{AccelComp}(a_{\cal{P}},v_{\cal{P}}, v_{\cal{E}}, gap)$
        \State $a_{\cal{E}}^{\scaleto{est}{3pt}} \gets
        \textit{RespEst}(v_{\cal{P}}, v_{\cal{E}}, gap)$
        \State $a_{\cal{E}} \gets \textit{Plausibility}(a_{\cal{E}}^{\scaleto{est}{3pt}} ,a_{\cal{E}}^{\scaleto{C}{3.0pt}}, v_{\cal{P}}, gap, D_{\cal{P}}^{\scaleto{max}{2.5pt}}) $

    \EndIf
    \State \textbf{return} $a_{\cal{E}}$
\EndProcedure

\Procedure{Plausibility}{$a_{\cal{E}}^{\scaleto{est}{3pt}} ,a_{\cal{E}}^{\scaleto{C}{3.0pt}}, v_{\cal{P}}, gap, D_{\cal{P}}^{\scaleto{max}{2.5pt}} $}
    \setstretch{1.2}
         \State $t_{\scaleto{gap}{3pt}}^{\scaleto{est}{3pt}},t_{\scaleto{gap}{3pt}}^{\scaleto{C}{3.0pt}}
         \gets \textit{GetTGap}(a_{\cal{E}}^{\scaleto{est}{3pt}} ,a_{\cal{E}}^{\scaleto{C}{3.0pt}}, v_{\cal{P}}, gap, D_{\cal{P}}^{\scaleto{max}{2.5pt}}) $ 
         \If{$t_{\scaleto{gap}{3pt}}^{\scaleto{C}{3.0pt}}>T_{\scaleto{gap}{3pt}}^{\scaleto{C}{3.0pt}} \And t_{\scaleto{gap}{3pt}}^{\scaleto{C}{3.0pt}}< t_{\scaleto{gap}{3pt}}^{\scaleto{est}{3pt}} \And t_{\scaleto{gap}{3pt}}^{\scaleto{C}{3.0pt}}<T_{\scaleto{gap}{3pt}}^{\scaleto{A}{3.0pt}}$}
          \State $a_{\cal{E}} \gets a_{\cal{E}}^{\scaleto{C}{3.0pt}} \textbf{\textit{ corrected CACC output applied}} $
         \ElsIf{$t_{\scaleto{gap}{3pt}}^{\scaleto{est}{3pt}}>T_{\scaleto{gap}{3pt}}^{\scaleto{C}{3.0pt}} \And  t_{\scaleto{gap}{3pt}}^{\scaleto{est}{3.0pt}}<T_{\scaleto{gap}{3pt}}^{\scaleto{A}{3.0pt}}$}
         \State $a_{\cal{E}} \gets a_{\cal{E}}^{\scaleto{est}{3pt}} 
         \textbf{\textit{ Response Estimator output applied}}$
         \Else{}
         \setstretch{1.0}
         \State $a_{\cal{E}} \gets
         a_{\cal{E}}^{\scaleto{A}{3.0pt}} 
         \textbf{\textit{ degrade to ACC}}$
         \EndIf
         \setstretch{1.0}
         \State \textbf{return} $a_{\cal{E}}$
\EndProcedure

\end{algorithmic}
\end{algorithm}

\subsection{Mitigator}

For each anomaly captured by the detector, Mitigator computes an alternate response overriding the  CACC controller response $a_{\cal{E}}$, to neutralize any potential adversarial impact.  Mitigator comprises the following components.
\begin{enumerate}
\item {\bf Response Estimator} is a pre-trained machine learning model analogous to Predictor, that generates an estimated acceleration $a_{\cal{E}}^{\mbox{est}}$. However, unlike Predictor (and indeed, \proc{COMP}), it uses only trusted sensory inputs, \eg, relative velocity and position of ${\cal{E}}$ and ${\cal{P}}$.
\item {\bf Plausibility Checker}  ensures that Response Estimator's output does  not compromise the  safety of ${\cal{E}}$, even under attack.
\end{enumerate}

\noindent  Algorithm \ref{alg:Mitigation} describes the  Mitigator functionality.  In the absence of anomaly, sensory inputs are typically sampled at a lower rate $F_{\mbox{normal}}$.  When Mitigator is invoked to handle an anomaly (lines $7$ through $10)$, the sensor sampling frequency is switched to a higher value $F_{\mbox{max}}$ to generate more accurate sensory data.  The  $a_{\cal{P}}$ received as anomalous message, and $a_{\cal{E}}$ computed using that value, are discarded. Instead, $a_{\cal{E}}$ is calculated approximately using the rate of change in the velocity of the ${\cal{P}}$ from the previous time step.  
%Response Estimator is invoked to output  $a_{\cal{E}}^{\mbox{est}}$ followed by plausibility checker.  
Lines $14$ through $21$ describe the plausibility checker functionality; it accounts for the worst case for safety, \eg, sudden halt of ${\cal{P}}$.  The resultant $t_{\mbox{gap}}$  is computed  for the scenario where $a_{\cal{E}}^{\mbox{est}}$ and corrected $a_{\cal{E}}$ were  applied.  The plausibility checker then determines the optimal choice out of  $a_{\cal{E}}^{\mbox{est}}$ and the corrected $a_{\cal{E}}$ that is both safe and efficient. If it fails to find such a value, the system falls back to  conservative ACC.  Consequently,  THW never reaches value less than minimum safe threshold $T_{\mbox{gap}}$. 
%while ma maintained at an optimal value. 

\subsection{Data Collector}

The Data Collector collects on-road driving data, which is aggregated and periodically communicated to the cloud for improving the ML models (see below).  The collected data includes (1) inputs to the CACC controller, \eg, preceding vehicle acceleration, inter-vehicle space headway, and the velocities of the two vehicles; (2) the acceleration value computed by the \proc{COMP} module of CACC in response to these inputs; and (3) an ``anomaly flag'' to indicate whether the response is classified as an anomaly by RACCON.

\subsection{Off-line Cloud Infrastructure}
The ML components of RACCON (Predictor and Response Estimator) are trained offline on trusted cloud servers and updated periodically, as new on-road CACC data is made available from the Data Collector modules of different vehicles subscribing to the RACCON service.  We assume that these communications cannot be corrupted.  This is viable in practice since we do not require real-time communication with the cloud.  Data can be transferred from the vehicles when a trusted connection to the cloud is available. 
%Correspondingly, the optimal anomaly threshold for each driving environment is also re-computed periodically.  
RACCON-enabled vehicles securely download the latest instances of trained Predictor and Response Estimators along with a list of anomaly thresholds for different driving environments, prior to CACC engagement in untrusted operating conditions.   

\section{RACCON Evaluation, Setup, and Attack Orchestration Methodology}
\label{sec:ExpSetup}

A unique aspect of our work is the extensive  experimental evaluation of RACCON.  In addition to showing the viability of RACCON itself, we believe our experiments provide a roadmap for evaluation of resiliency in other  connected vehicular applications as well.  

\subsection{Data Generation}
\label{subsec:DataGen}

A key challenge with evaluating ML-based solutions is the need for realistic data.  As discussed in Section \ref{sec:DesignConst}, we avoid the need for real vulnerability data by training the ML components to learn NBM (for which there is plentiful data \textit{on deployment}).  However, our {\em experimental evaluation} obviously needs to be done  before deployment, when in-field data is not yet available.  Consequently, we generate normal driving data using a state-of-the-art physical automotive research simulator, RDS1000\textregistered\ \cite{rdswebsite} and a software system replicating a representative CACC controller functionality (described in \ref{sec:Background}).   Data collected from the  simulator is fed to the CACC software system to generate  vehicular trajectories. RACCON detection and mitigation components are  integrated with the CACC system.  Attacks are orchestrated by manipulating the inputs to the RACCON-integrated CACC system.  The impacts of the attack (and our mitigation) are computed by modifying  vehicular parameters (\eg, acceleration of the ego vehicle, THW between vehicles, etc.).  

We curated a dataset corresponding to vehicles engaged in CACC, operating in $24$ different driving environments.  These environments were programmed as a cross-product of the following parameters: (i) Road terrain (highway, suburban and urban); (ii) Weather (clear, windy, snowy, rainy); and (iii) Time of day (day, night).  The set of parameters (terrain, weather, and time of day) are typically used to analyze driving patterns \cite{nhtsa}.  We also added ambient traffic to obtain realistic vehicular trajectory data.  Each of the $24$ datasets corresponds to about $15$ minutes of driving time and constitutes approximately $90,000$ samples collected at a frequency of $100$Hz.  The data collected provides the preceding vehicle trajectory; ego vehicle response is computed using the \proc{COMP} controller from Section \ref{subsec:cacc}.  The global dataset is generated by aggregating data from all environments, and is split $80$-$20$ into training and test data.

\subsection{Attack Taxonomy}
\label{subsec:Taxonomy}
\begin{figure}
\centering
\includegraphics[width= 0.85\columnwidth]{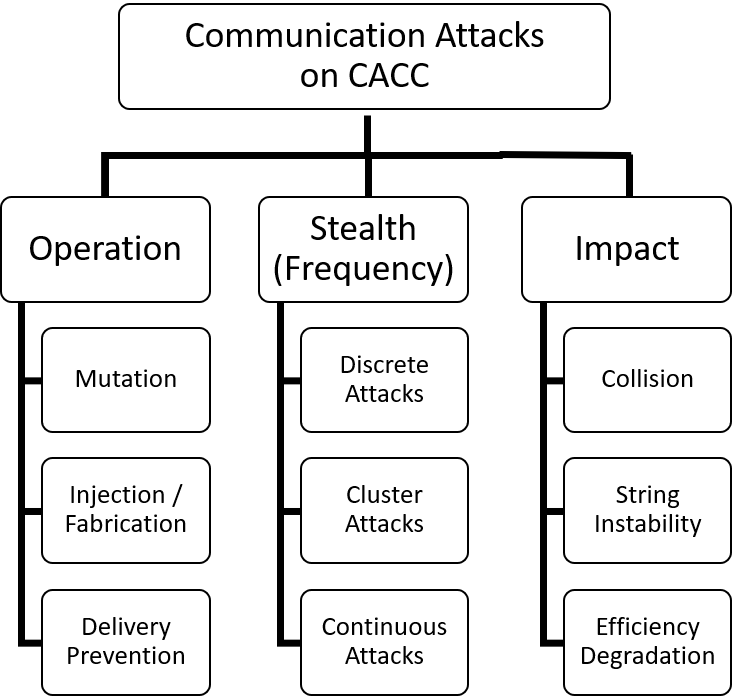}
\caption{{\small Taxonomy of Communication Attacks on CACC}}
\label{fig:taxonomy}
\end{figure}

%Recall that the goal of RACCON is to provide resiliency against the spectrum of V2V attacks.  
Since the security paradigm of V2V communications is continuously increasing in complexity, it is challenging to develop an evaluation strategy to comprehensively cover the attack space.  All previous works on V2X attack detection only focused on specific attack instances, \eg, Biron {\em et al.} \cite{CACCFaultDetection} only target jamming and flooding attacks, and  Jagielski {\em et al.} \cite{jagielski2018threat} focus on specific mutation attacks. Such evaluation does not provide adequate evidence of resiliency against other potentially unknown attacks.   

\begin{table*}
\centering
\caption{{\small Representative N-day Attack Instances. All relevant combinations of the operation, frequency and impact features for each attack mechanism indicated by ``\cmark'' }}
\label{tbl:AttackMech}
\resizebox{\textwidth}{!}{%
\begin{tabular}{cccccccccccc}
\toprule
\multirow{2}{*}{\shortstack{ \\ Attack \\ Mechanism }} & \multicolumn{2}{c}{ Attack Origin } & \multicolumn{3}{c}{ Operation } & \multicolumn{3}{c}{ Frequency } & \multicolumn{3}{c}{ Impact } \\
\cmidrule(lr){2-3}
\cmidrule(lr){4-6}
\cmidrule(lr){7-9}
\cmidrule(lr){10-12}
& Preceding Vehicle & MITM & Mutation & Fabrication & Delivery Prevention & Discrete & Cluster & Continuous & Collision & Efficiency degradation & String Instability \\
\midrule
Message falsification & \cmark & \cmark & \cmark & & & \cmark & \cmark & \cmark & \cmark & \cmark & \cmark \\
%DoS & \cmark & \cmark & & \cmark & \cmark & \cmark & \cmark & \cmark & \cmark & \cmark & \cmark \\
DoS (Jamming) & & \cmark & & & \cmark & \cmark & \cmark & & & \cmark & \cmark \\
DoS (Flooding) & \cmark & \cmark & & \cmark &\cmark & \cmark & \cmark & \cmark & \cmark & \cmark & \cmark \\
Masquerade & & \cmark & \cmark & \cmark & & \cmark & \cmark & \cmark & \cmark & \cmark & \cmark \\
Replay & & \cmark & & \cmark & & \cmark & \cmark & & \cmark & \cmark & \cmark \\
Misdirection & & \cmark & & & \cmark & \cmark & \cmark & & & \cmark & \cmark \\
\bottomrule
\end{tabular}
}
\end{table*}

We address this problem by developing a comprehensive taxonomy of V2V attacks on CACC (Fig.~\ref{fig:taxonomy}) that is  used to systematically navigate the attack space. The taxonomy is inspired by threat modeling approaches in hardware and system security \cite{pieee}, but adapted for V2V adversaries.  The idea is to represent a V2V attack through three features, \viz, stealth, operation, and impact.   This feature combination forms a holistic characterization of any attack under the  RACCON adversary model.    In particular, since the adversary  is confined to V2V communications, the only choices for the adversary are to (1) mutate an existing message, (2) fabricate a new message, and (3) prevent the delivery of a message.  Correspondingly, since the message payload constitutes the preceding vehicle's acceleration, the impact of an attack can be to (1) increase the probability of collision (by reporting a lower than actual acceleration value), (2) reduce efficiency through an increased headway (by reporting a higher than actual acceleration value), or (3) creating instability (\eg, through random mutation of the actual value).  We refer to deviations by a positive bias as {\em collision attacks}  and deviations by a negative bias  as {\em efficiency degradation attacks}.   Note that the taxonomy is oblivious to the  \textit{mechanics} of the attack (\eg, man-in-the-middle, rogue vehicle, hardware-software modules of the ego vehicle, etc.), but only considers the effect on V2V messages. For instance, delivery prevention operation accounts for jamming, flooding, channel subversion, etc., each of which can be carried out through a variety of ways.  Table \ref{tbl:AttackMech} shows how the taxonomy accounts for  different well-known attacks.  The focus on {\em attack characteristics} rather than the mechanics enables the taxonomy to provide a comprehensive classification of V2V attacks.    
%Correspondingly, the use of the taxonomy to systematically create V2V attacks ensures that the attack spectrum has been comprehensively explored in evaluation, and consequently provide assurance on RACCON resliency.  
%We consider the development of V2V taxonomy and its use for attack generation to be a key step towards a systematic evaluation methodology for connected autonomous vehicular applications.

\subsection{Attack Orchestration Methodology}
\label{subsec:attacks}

We used the taxonomy above to develop a systematic attack orchestration framework.  Attacks are represented as $3$-tuples, representing the three features identified in the taxonomy.  Delivery prevention attacks are realized through intermittent or absent communication.  Mutation and fabrication attacks are realized through fake acceleration messages that deviate from ground truth. We consider four different ways for generating fake accelerations:

\begin{gather}
    a_{\cal{P}}^{\scaleto{fake}{6pt}} = a_{\cal{P}}^{\scaleto{true}{4.5pt}} \pm b \\
    a_{\cal{P}}^{\scaleto{fake}{6pt}} = a_{\cal{P}}^{\scaleto{true}{4.5pt}} \pm bt \\
    a_{\cal{P}}^{\scaleto{fake}{6pt}} = a_{\cal{P}}^{\scaleto{true}{4.5pt}} \pm bsin(ft)\\
    a_{\cal{P}}^{\scaleto{fake}{6pt}} = a_{\cal{P}}^{\scaleto{true}{4.5pt}} \pm random
\end{gather}

Equation (1) represents a constant bias added to the ground truth. Equations (2) and (3) represent  linear and  sinusoidal time-varying biases, respectively. Given a specific combination of attack features (\eg, discrete mutation attack with collision as targeted impact), the framework permits attack impact simulation.  We use THW ($t_{\mbox{gap}}$) as a natural measure to quantify the risk of collision or the extent of efficiency degradation.  An erratic change in $t_{\mbox{gap}}$ can also potentially indicate string instability in the traffic.

\subsection{Summary of Experiments}

Evaluation of CAV application resiliency must address a variety of orthogonal facets. Note that within the broad umbrella of ML-based resiliency, the number of architectural parameters available for a security designer to tweak is dauntingly large. This includes the choice of ML model, anomaly threshold, adversary classifications, etc.  In addition to evaluating the quality of infrastructure, the methodology must enable systematic estimation of these parameters.    Following is an overview of the experiments performed to evaluate RACCON.  We elaborate on the experiments in Sections \ref{sec:DataVal} through \ref{sec:DetectorSubversion}.

\begin{enumerate}
\item {\bf Data Validation:} For our conclusions to be meaningful, it is critical that the data we use is realistic.  We validate that the vehicular driving patterns reflected in our simulation data conform to real-world patterns from a public dataset. (Section \ref{sec:DataVal})
% with public dataset that the pattern of vehicular behavior reflected in our simulation data indeed conforms to real-world patterns.  (Section \ref{sec:DataVal})
\item {\bf Identification of Appropriate ML Model:} Implementing Predictor and Response Estimator functionalities requires selecting and tuning the appropriate ML architecture.  We develop a systematic evaluation methodology to address this problem. (Section \ref{sec:OfflineML})
\item{\bf Attack Impact Analysis:} The viability of attack orchestration framework for RACCON evaluation depends on the quality of the orchestrated attacks themselves.  We develop a methodology to analyze attacks, in terms of stealth and impact. (Section \ref{sec:Impact})
%This analysis enables selection of appropriate attacks to evaluate RACCON resiliency.
\item {\bf Anomaly Detection Threshold:} A key factor in the effectiveness of RACCON is the identification of {\em anomaly threshold}, \ie, the extent of deviation from normal behavior pattern that would be classified as a potential threat. Selecting an appropriate threshold involves balancing the trade-off between maximizing attack detection accuracy and minimizing false alarms. %Selecting an appropriate threshold entails comprehending and balancing the needs of ensuring that an attack does not go undetected and minimizing the number of false alarms.  
We present a series of experiments to achieve this balance. (Section \ref{sec:AnomalyThresh})
\item {\bf V2V Attack Resilience:}  The central component of our evaluation shows the robustness of RACCON against various V2V attacks.  (Section \ref{sec:ResiliencyEval})
\item {\bf Resilience Against Detector Subversion:} Since RACCON depends on  ML-based predictions, it is exposed to  adversaries that aim to subvert the ML component.  We call such adversaries {\em Detection Subversion Adversaries}, and evaluate the resiliency of RACCON against them.  We also present an interesting connection between anomaly threshold and detection subversion. (Section \ref{sec:DetectorSubversion})
\end{enumerate}

\section{Data Validation}
\label{sec:DataVal}

A key challenge with using simulator data is to ensure that it  is realistic.  Unfortunately, there is no available repository of sufficient real-world driving data across different driving scenarios. Indeed, the lack of available real-world data is the reason why we rely on simulated data in the first place.  To address this problem, we observe that while sustained data over a period of time is unavailable, there are datasets that provide short-duration driving patterns. These snippets can then be used to corroborate data obtained from the simulator under similar driving conditions.    
%Consequently, one can use these  snippets to check if they show the similar driving trend under the similar environmental scenario as data obtained from the simulator.

\begin{figure}
\centering
\includegraphics[width=\columnwidth]{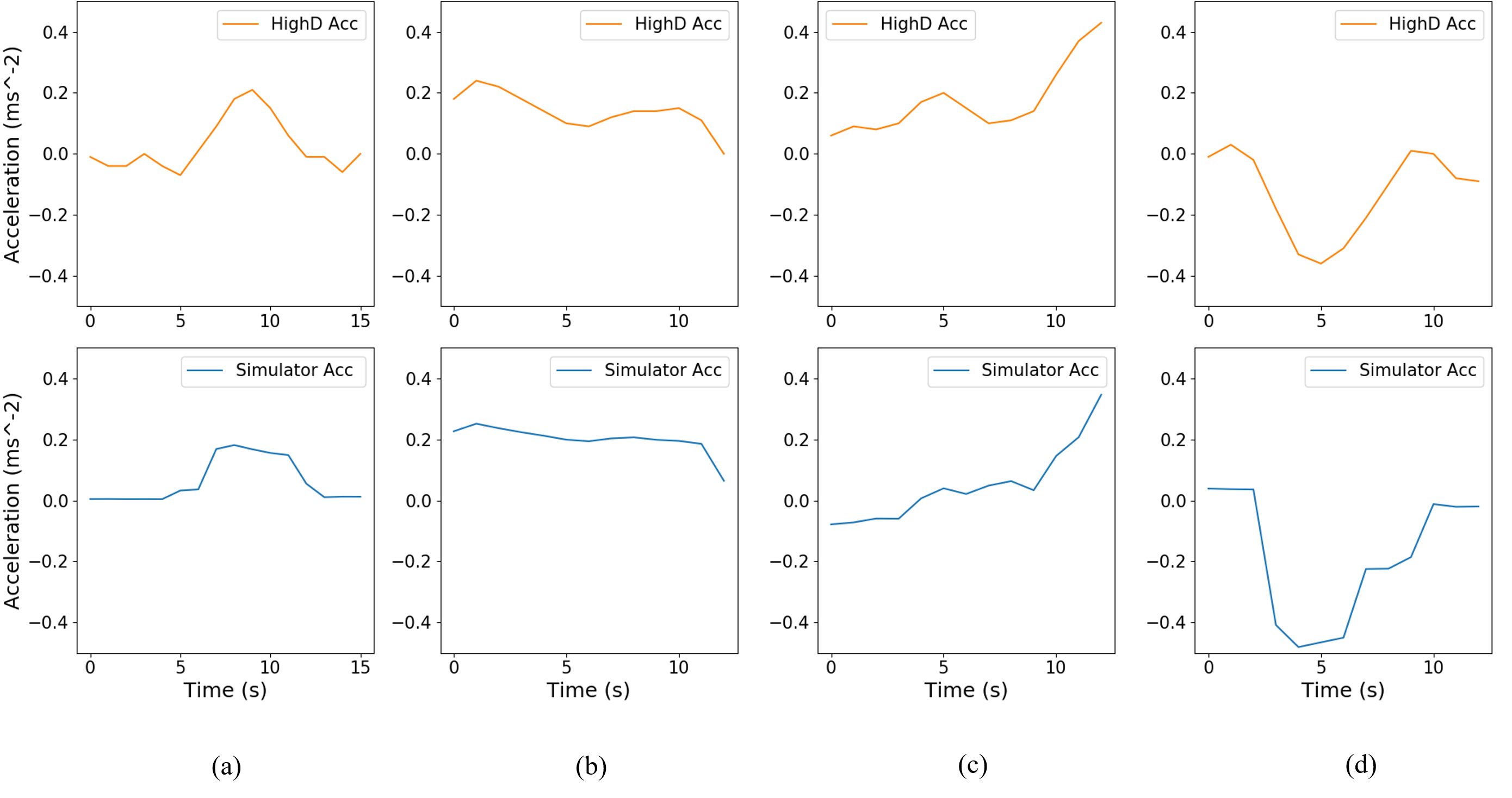}
\caption{{\small Correlation Between Simulated Data and HighD. Each plot indicates correlation between the acceleration trajectory of an arbitrary vehicle in HighD and the simulated vehicle.}}
\label{fig:data_val}
\end{figure}

We carried out this experiment with HighD dataset \cite{highDdataset} that provides trajectory data corresponding to real vehicles driving in German highways.  The length of individual vehicle trajectories is approximately $15$ seconds.  We compare acceleration patterns of similar length trajectories collected from the simulator.  Fig.~\ref{fig:data_val} shows sample comparisons for four vehicles from HighD data.  The results clearly indicate that the acceleration patterns from the simulator correlate closely with HighD data.

%\subsubsection{Data Generation}
%We curated a diverse global dataset that corresponds to vehicles engaged in CACC, operating in $24$ different driving environments. We simulated these environments on RDS1000 physical simulator considering a combination of the following parameters: (i) Road terrain (Highway, Suburban and Urban); (ii) Weather (Clear, Windy, Snowy, Rainy); and (iii) Time of day (Day, Night). This parameter set is standard and used in many contexts to define factors influencing driving patterns \cite{nhtsa}. In addition to the creating various driving environments, we also add ambient traffic in the simulation to obtain realistic vehicular trajectory data. Each dataset pertaining to a driving environment, corresponds to about $15$ mins of driving time and constitutes approximately $90,000$ samples collected at a frequency of $100$Hz. The data collected from each environment provides the preceding vehicle trajectory. Subsequently, a software model of the CACC acceleration computation module generates the response of the following ego vehicle. In this manner, data corresponding to a CACC engagement is prepared for each driving environment. Ultimately, a global dataset is generated by cascading data from all the CACC engagements and is then split into training and test data.  

%\subsubsection{Data Validation}

%A key challenge in validating the simulation data is the unavailability of realistic

%We validate the data collected from the RDS1000 simulator by comparing it with an existing benchmark dataset, HighD \cite{highDdataset}, that provides trajectory data corresponding to real vehicles driving in German highways. The length of individual vehicle trajectories is in typically within $15$ seconds in the dataset. We compare acceleration patterns of similar length trajectories collected from the simulator with HighD trajectories. Results show that the acceleration patterns from the simulator vehicular data correlate closely with HighD data. Figure \ref{fig:data_val} shows comparison between sample trajectories of four vehicles from HighD data and the trajectories generated with the simulator. 

\section{Fidelity of ML Models}
\label{sec:OfflineML}

Viability of RACCON critically depends on the presumption that ML Models involved (Predictor and Response Estimator) have high accuracy. We can formulate the ML regression problem in two ways: (i)stateless prediction and (ii)time-series prediction. Cumulatively, these result in a prohibitively large space ML architecture choices. It is important to navigate this space systematically and converge to an optimal architecture. The ML model must address two orthogonal requirements: (1) avoid false alarms for benign messages and (2) accurately classify malicious messages as anomalous.   Furthermore, it must be possible to perform real-time prediction under the computation and storage constraints of automotive systems.  Finally, since driving patterns vary according to driving conditions, we must determine whether each driving environment requires a  customized ML model.  
%Our experiments in this section answer those crucial questions.

\subsection{Identifying ML Architecture}

\begin{figure*}
\centering
\subfloat[]{
  \includegraphics[width= 0.48\textwidth]{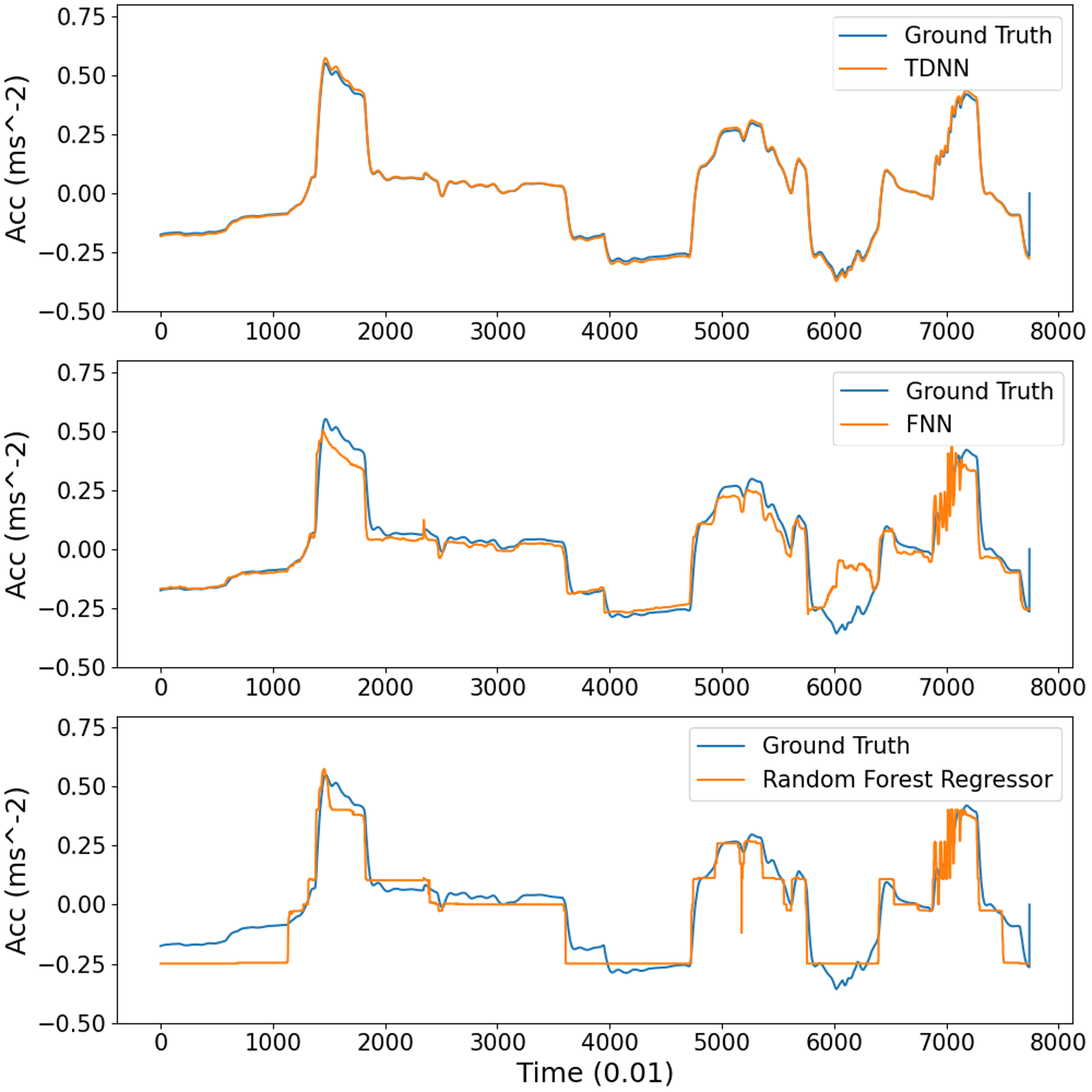} 
}
\subfloat[]{
  \includegraphics[width= 0.48\textwidth]{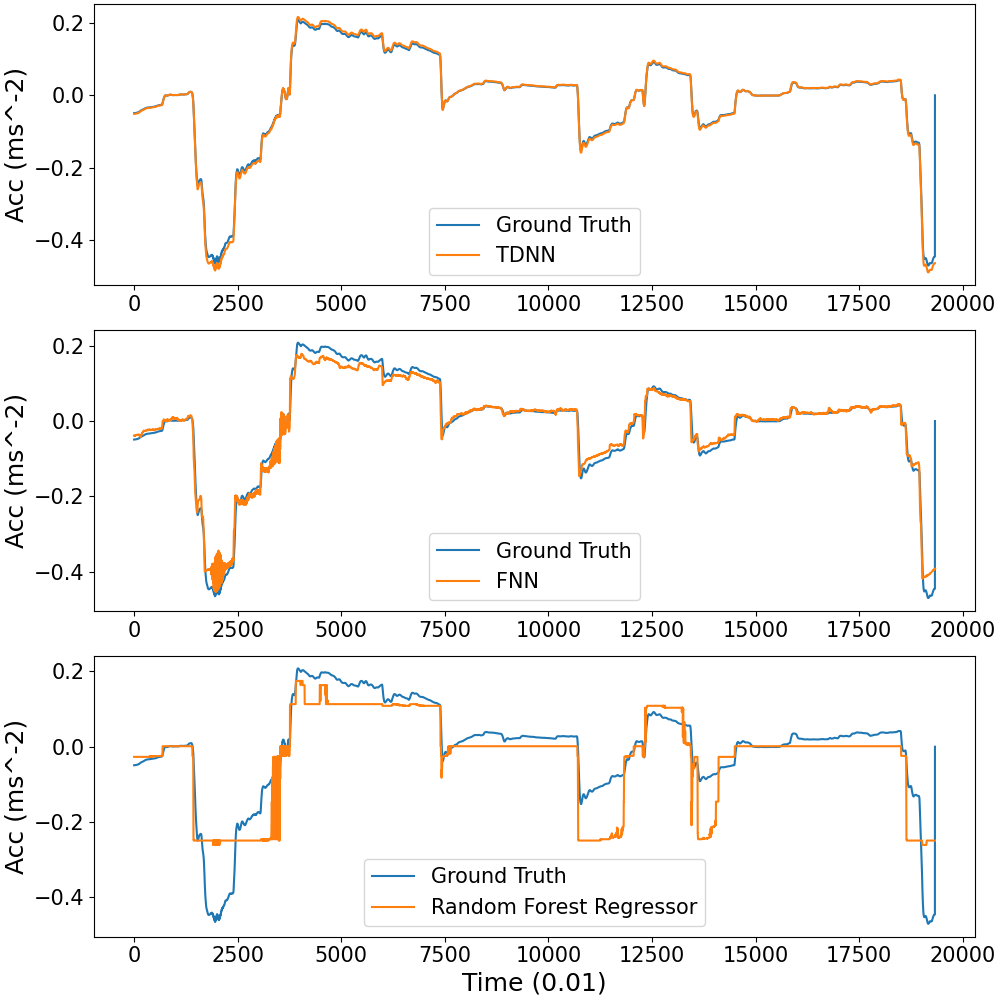} 
}

\caption{{\small Prediction of TDNN, FNN and Random Forest in Benign  Environments.  (a) Highway-Day-Windy.  (b) City-Night-Snow.}}
\label{fig:fnn_vs_others}
\end{figure*}

Since detecting malicious activity essentially involves identifying anomalous behavior, it is imperative that the model learns NBM (\ie, estimating the normal behavior of CACC controller) accurately for effective performance in adversarial settings. Furthermore, {\em efficiency} of a resiliency solution depends primarily on the prediction accuracy under benign scenarios, since most of the messages encountered by vehicles in field are likely benign. Our methodology entails the following steps to determine the appropriate ML architecture.
\begin{enumerate}
    \item Find a set of candidate architectures that can satisfy automotive resource constraints.
    \item Discard candidates that do not provide acceptable prediction accuracy under benign conditions.
    \item Of the remaining candidates, select the architecture with highest accuracy under malicious conditions.
\end{enumerate}
In our evaluations, we started with five architectures: Random Forest Regressor (RF), Support Vector Machine (SVM), and  Feed-forward Neural Network (FNN) are examined for stateless prediction;  Univariate Time Delayed Neural Network (TDNN) and Multivariate Long Short-Term Memory (LSTM) network are examined for time-series prediction.  Architectures more sophisticated than LSTM were estimated to be too complex, given the  constraints of automotive systems.   For these candidates, we apply a two-step triage process based on prediction accuracy in benign environment.  In the first step,  we compute the Mean Absolute Error (MAE) in prediction, under six different driving environments, for each ML architecture. This provides a ``coarse'' evaluation of accuracy and facilitates identification of a small subset of candidates  (Table \ref{tbl:MLModels} ).  Clearly RF, TDNN, and FNN show much better accuracy than SVM and LSTM. In the next step, we examine them more closely to identify any local ``kinks''.  Fig.~\ref{fig:fnn_vs_others} plots the accuracy of Predictor in two different environments. Note that RF is ineffective in capturing minute variations in acceleration (indicated by several flat lines in prediction). This behavior can be attributed to the fact that the RF regressor ignores  minute variations in the data as noise. Since tracking such variations is critical for accurate anomaly detection, RF is discarded as a viable candidate.

%In-filed resiliency of RACCON relies heavily on the accuracy of the machine learning components: predictor and Response Estimator. The quality of training data and the appropriate choice of the machine learning architectures for Predictor and Response Estimator determine the accuracy of the models. In this section we present our experimental analysis for determining the optimal ML architectures used for Predictor and Response Estimator. 

%%\subsection{ML Architectures}
%Predictor and Response Estimator ML models act as intelligent function approximators of the underlying CACC acceleration computation module. As we highlighted in Section \ref{sec:Background} the regression problem is formulated in two ways. (1) Stateless Prediction: Predict the acceleration value based on the current states of the inputs of a CACC acceleration computation module. (2) Time-series Prediction: Predict the acceleration from the current values and the previous history of acceleration and/or other inputs. Different machine learning architectures are considered for each approach. Random forest regressor, support vector machine (SVM) and a feed-forward neural network (FNN) are examined for stateless prediction problem. Univariate time-delayed neural network (TDNN) and multi-variate long short-term memory (LSTM) network are examined for the time-series prediction problem. The architecture that provides the best detection accuracy and effective mitigation, with the least amount of computation and storage overhead is considered for Predictor and Response Estimator implementation.

\begin{table}
\centering
\caption{{\small Mean Absolute Error in the prediction of ego vehicle acceleration under six different test driving environments}}
\label{tbl:MLModels}
\resizebox{\columnwidth}{!}{%
\begin{tabular}{cccccc}
\toprule
\multirow{2}{*}{ \shortstack{Test\\ Environment} } & \multicolumn{5}{c}{ ML Model } \\
\cmidrule{2-6}
& RF & FNN & LSTM & TDNN & SVM \\
\midrule
Env 1 & 0.040 & 0.021 & 0.155 & 0.007 & 0.440 \\
Env 2 & 0.149 & 0.177 & 1.640 & 0.027 & 1.057 \\
Env 3 & 0.101 & 0.116 & 0.985 & 0.021 & 0.787 \\
Env 4 & 0.075 & 0.089 & 1.166 & 0.010 & 0.510 \\
Env 5 & 0.199 & 0.310 & 1.130 & 0.035 & 0.364 \\
Env 6 & 0.062 & 0.073 & 0.201 & 0.010 & 0.987 \\
\bottomrule
\end{tabular} 
}
\end{table}

%Table \ref{tbl:MLModels} shows predictor models considered. Test sets corresponding to five arbitrary driving scenarios are considered and the mean absolute errors are recorded. It can be seen that the the univariate time delayed neural network (TDNN), feed-forward neural network (FNN) and random forest regressor (RF) show better accuracy than the rest, in normal benign operating conditions. 

% \begin{figure*}
% \centering
% \includegraphics[width= 0.95\textwidth]{FNN_vs_others.png}
% \caption{{\small Predictor estimate of the CACC controller output for different ML architectures: Each of the two columns corresponds to a driving environment. Top subplots show the TDNN estimate. Middle subplots show FNN estimate and the bottom subplots show RF Regressor estimate }}
% \label{fig:fnn_vs_others}
% \end{figure*}

%Figure \ref{fig:fnn_vs_others} shows the accuracy of Predictor in two different test driving environments. Random forest regressor seems to be ineffective in capturing minute variations in the acceleration as indicated by several flat lines in the prediction. This behavior could be attributed to the fact that the RF regressor might have been trained to ignore the minute variations as noise in the data. However, tracking such variations is critical for accurate anomaly detection. Hence, RF regressor is not considered as a potential architecture for the ML systems in RACCON.

FNN and TDNN are further examined under simulated attacks to determine anomaly detection and mitigation efficacy.  In each attack, malicious acceleration values are generated by adding a bias (constant or sinusoidal) to the ground truth.  Clearly, FNN performs significantly better than TDNN in mitigating attacks, as indicated by the resultant THW values in Table \ref{tbl:FNNvsTDNN} \footnote{We believe the better performance of FNN over TDNN is due to the stateless design of the CACC controller. The stateless FNN model captures the context well and approximates the controller behavior while time-dependant regression models learn spurious temporal dependencies making them ineffective in detecting anomalous inputs.}  \textit{Based on these results, FNN is determined as the appropriate ML architecture for the RACCON detection system.}.

%%FNN and TDNN architectures are further examined under the context of simulated attacks to determine the accuracy of anomaly detection and mitigation effectiveness. In each attack, the malicious acceleration values are generated by adding a bias (constant or sinusoidal) to the ground truth. Table
%%\ref{tbl:FNNvsTDNN} shows the resultant mitigation of FNN-based RACCON and TDNN-based RACCON as opposed to a naive CACC system with no resiliency. It can be seen that, TDNN is not very effective in mitigating attacks in several cases. We believe the better performance of FNN over others could be attributed to the inherently stateless design of the underlying CACC controller. Therefore a stateless FNN model is able to better approximate the CACC behavior in normal conditions and capture anomalies effectively, as opposed to time-dependant regression models.   

\begin{table*}
\centering
\caption{{\small Resultant THW for TDNN and FNN predictors under four different attacks}}
\label{tbl:FNNvsTDNN}
\resizebox{\textwidth}{!}{%
\begin{tabular}{ccccccccccccc}
\toprule
\multirow{2}{*}{\shortstack{ \\Time \\ Headway }}&  \multicolumn{3}{c}{Cluster Attack (Bias:1.5) } & \multicolumn{3}{c}{ Cluster Attack (Bias:-0.8) } & \multicolumn{3}{c}{ Continuous Attack (Bias:0.1) } &
\multicolumn{3}{c}{ Continuous Attack (Bias:sin(0.05t)) }\\
\cmidrule(lr){2-4}
\cmidrule(lr){5-7}
\cmidrule(lr){8-10}
\cmidrule(lr){11-13}
& FNN & TDNN & Naive CACC & FNN & TDNN & Naive CACC & FNN & TDNN & Naive CACC & FNN & TDNN & Naive CACC \\
\midrule
THW $<$ 0.55s & 0\% & 30.85\% & 80.64\% & 0\% & 0\% & 0\% & 0\% & 63.22\% & 63.22\% & 0\% & 20.47\% & 21.14\% \\
THW: $\{0.55-0.75s\}$ & 100\% & 65.81\% & 19.36\% & 100\% & 55.89\% & 34.24\% & 100\% & 36.78\% & 36.78\% & 100\% & 77.97\% & 78.86\% \\
THW $>$0.75s & 0\% & 3.34\% & 0\% & 0\% & 44.11\% & 65.76\% & 0\% & 0\% & 0\% & 0\% & 1.56\% & 0\% \\ 
\bottomrule
\end{tabular}  
}
\end{table*}

\subsection{Environment Specific Models vs Global Model}

We investigate whether one global Predictor model can provide sufficient resiliency or a unique model is necessary for each driving scenario.  Firstly, we train  a unique Predictor model for each driving scenario as well as a cumulative global model. We then compare the global model against all unique models in normal operating conditions to determine the optimal approach.  Table \ref{tbl:UniqueGlobal} shows the mean absolute error for using a global predictor model vis-a-vis unique models.  The error corresponding to the global model is generally less than (but typically close to) unique models. Even when the error is greater, \eg, for $\langle\mbox{suburban},\mbox{night},\mbox{snow}\rangle$, the difference is insignificant.  Consequently,  using a global predictor to estimate $a_{\cal{E}}^{pred}$ is sufficient, obviating the need for unique models for different driving scenarios.\footnote{The conclusion that the global model performs better than unique tailor-made models for some  driving conditions is somewhat surprising. One reason is that it usually has more training data, incorporating driving patterns from many individual scenarios, resulting in a better accuracy.  This holds true for practical deployments as well as our experimental setup.} 

%Since driving patterns vary according to the different driving conditions, we investigate whether different predictor models are necessary for different driving scenarios or one global model trained by curating data from all driving scenarios is sufficient. To address this question, we trained a unique predictor model for each driving scenario, and also a cumulative, global model. We compare the global model against all the unique models in normal operating conditions to determine the approach that gives better accuracy.
%Table~\ref{tbl:UniqueGlobal} shows the mean absolute error for using a global predictor model vis-a-vis unique models for each scenario.  We observe that the error of the global model is less than (but typically close to) unique models; even when the error is greater, \eg, for $\langle\mbox{suburban},\mbox{night},\mbox{snow}\rangle$ scenario, the difference is insignificant.  Consequently,  using a global predictor to compute $a_{\cal{E}}^{pred}$ is sufficient, obviating the need for unique models for different driving scenarios.\footnote{The reason is global model often performs better than unique, tailor-made models is that it usually has more training data, incorporating driving patterns from many individual scenarios, resulting in a better accuracy.  This holds for practical deployments as well as our experimental setup.}

\begin{table*}
\centering
\caption{{\small Mean absolute error in $24$ different driving environments for Global and Environment-specific Predictors}}
\label{tbl:UniqueGlobal}
\resizebox{\textwidth}{!}{%
\begin{tabular}{ccccccccccccccccc}
\toprule
\multirow{3}{*}{\shortstack{ \\ Road \\ Infrastructure }} & \multicolumn{8}{c}{ Day } & \multicolumn{8}{c}{ Night } \\
\cmidrule(lr){2-9}
\cmidrule(lr){10-17}
& \multicolumn{2}{c}{ Rain } & \multicolumn{2}{c}{ Snow } & \multicolumn{2}{c}{ Clear } & \multicolumn{2}{c}{ Windy } & \multicolumn{2}{c}{ Rain } & \multicolumn{2}{c}{ Snow } & \multicolumn{2}{c}{ Clear } & \multicolumn{2}{c}{ Windy } \\
\cmidrule(lr){2-3}
\cmidrule(lr){4-5}
\cmidrule(lr){6-7}
\cmidrule(lr){8-9}
\cmidrule(lr){10-11}
\cmidrule(lr){12-13}
\cmidrule(lr){14-15}
\cmidrule(lr){16-17}

& \ Unique & \ Global & \ Unique & \ Global & \ Unique & \ Global & \ Unique & \ Global & \ Unique & \ Global & \ Unique & \ Global & \ Unique & \ Global & \ Unique & \ Global \\
\midrule
Highway &0.070 &0.051 &0.130 &0.131 &0.045 &0.044 &0.059 &0.058 &0.082 &0.073 &0.087  &0.084 &0.071 &0.068 &0.098 &0.091 \\
Suburban &0.125 &0.119 &0.038 &0.141 &0.193 &0.178 &0.103 &0.121 &0.101 &0.140 &0.074 &0.096 &0.133 &0.119 &0.110 &0.212 \\
City &0.081 &0.056 &0.084 &0.051  &0.212 &0.050  &0.005 &0.016 &0.051 &0.140 &0.059 &0.038 &0.053 &0.052 &0.034 &0.042 \\
\bottomrule
\end{tabular}
}
\end{table*}

%%%%%%%%%%%%%%%%%%%%%%%%%%%%%%%%%%%%%%%%%%%%%%%%%%%%%%%%%%%%
%%%%%%%%%%%%%%%%%%%%%%%%%%%%%%%%%%%%%%%%%%%%%%%%%%%%%%%%%%%%

\section{Attack Impact Analysis}
\label{sec:Impact}

\begin{figure*}
\centering
\includegraphics[width= 0.95\textwidth]{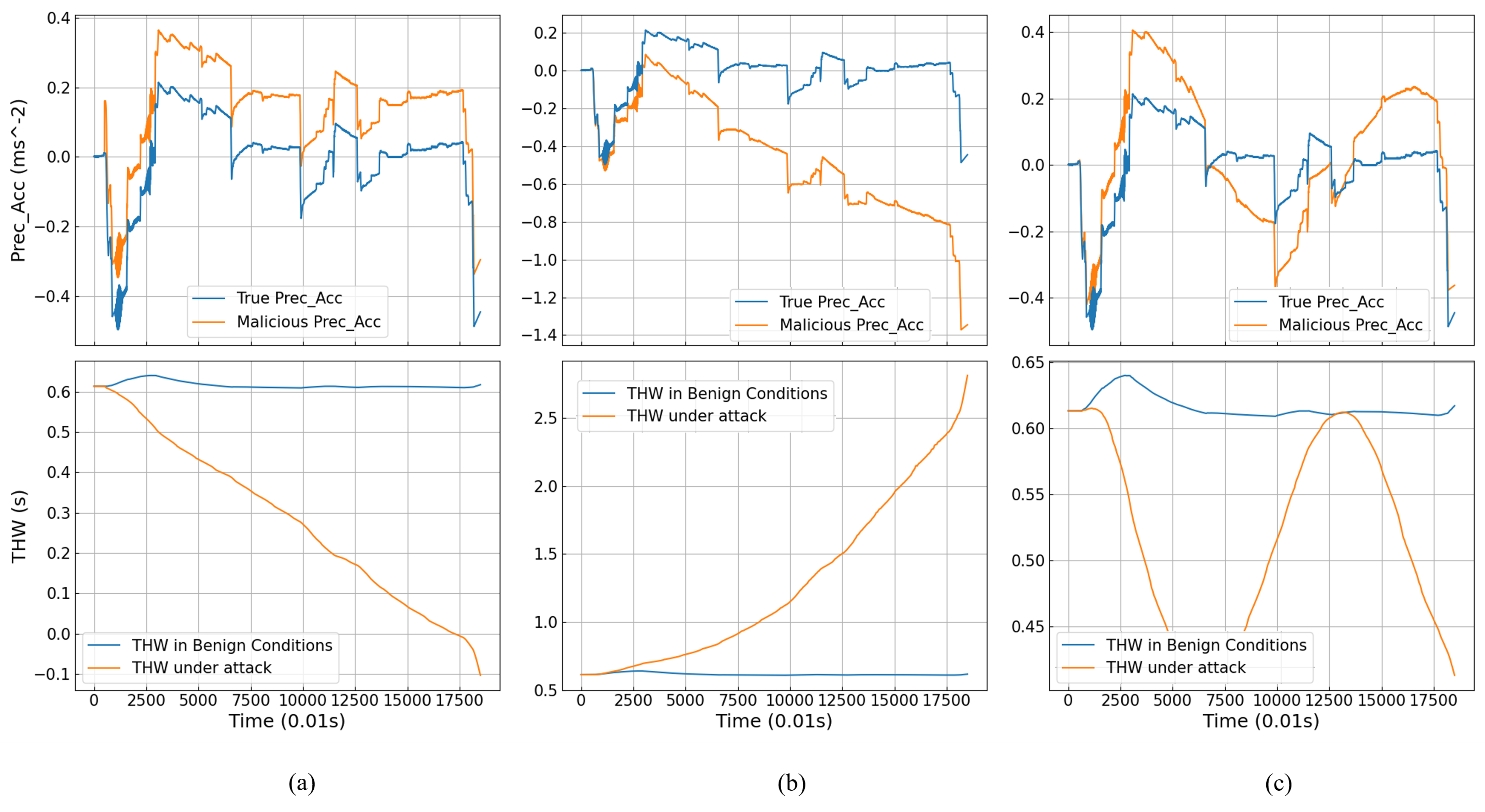}
\caption{{\small Impact of Continuous Attacks. (a) Constant Bias 0.15. (b) Linear Bias -0.005t. (c) Sinusoidal Bias 0.2sin(0.005t). }}
\label{fig:continuous_attacks}
\end{figure*}

\begin{figure*}
\centering
\includegraphics[width= 0.95\textwidth]{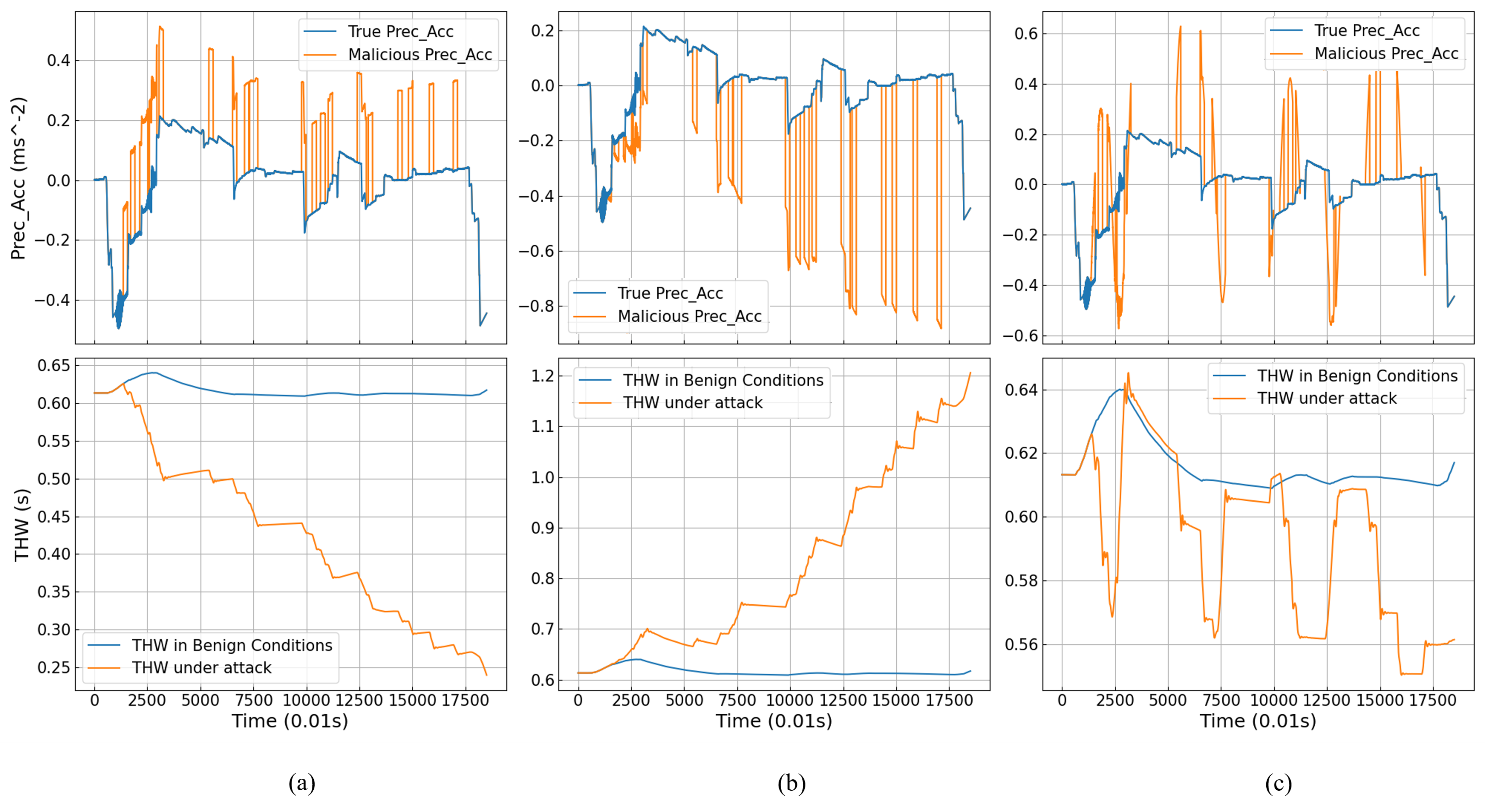}
\caption{{\small Impact of Cluster Attacks. (a) Constant Bias 0.3; (b) Linear Bias -0.02t. (c) Sinusoidal Bias 0.5sin(0.05t).}}
\label{fig:cluster_attacks}
\end{figure*}

\begin{figure*}
\centering
\includegraphics[width= 0.95\textwidth]{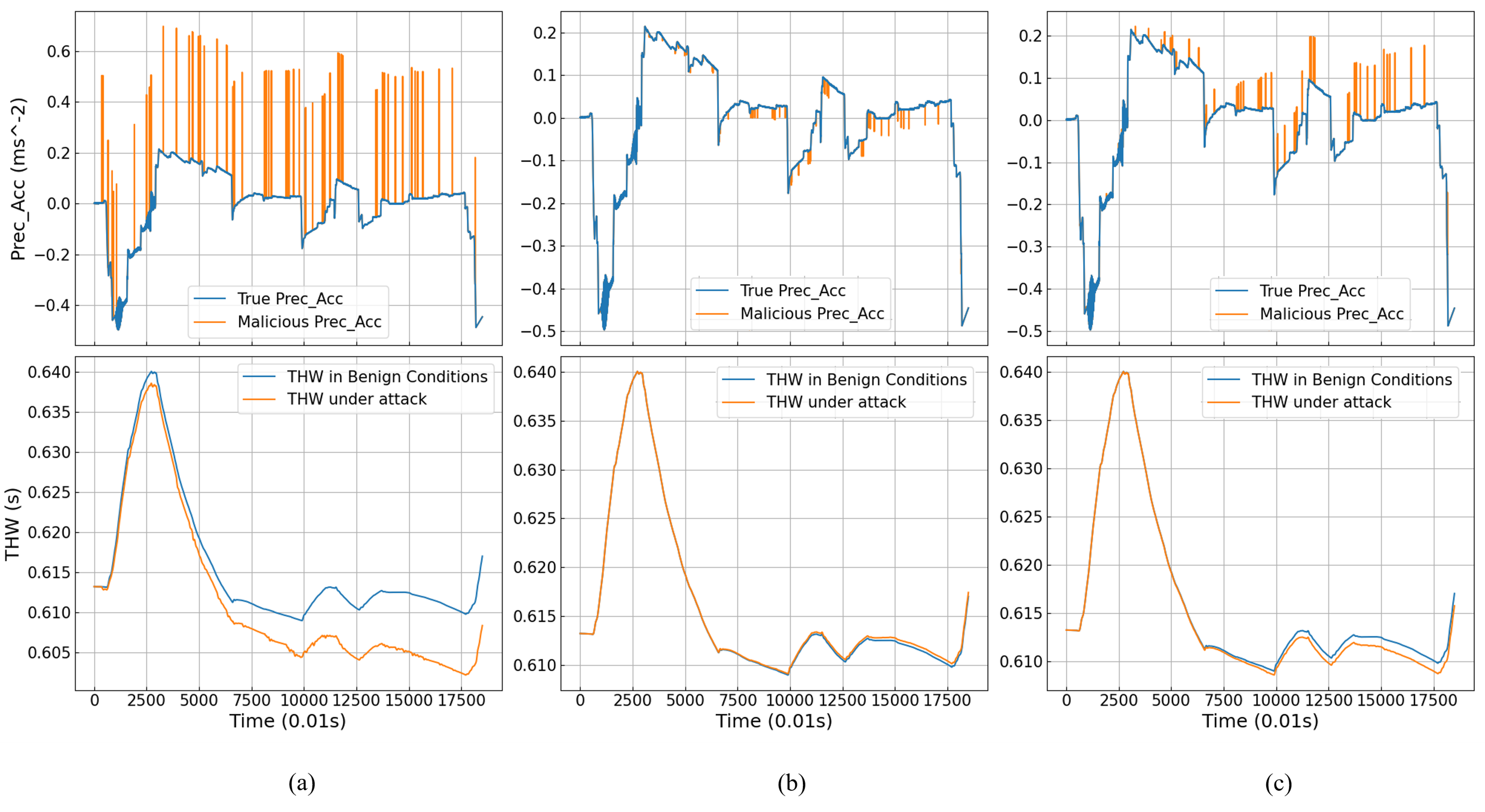}
\caption{{\small Impact of Discrete Attacks. (a) Constant Bias 0.5. (b) Linear Bias -0.005t. (c) Sinusoidal Bias 5sin(0.005t).}}
\label{fig:discrete_attacks}
\end{figure*}

\begin{figure*}
\centering
\includegraphics[width= 0.95\textwidth]{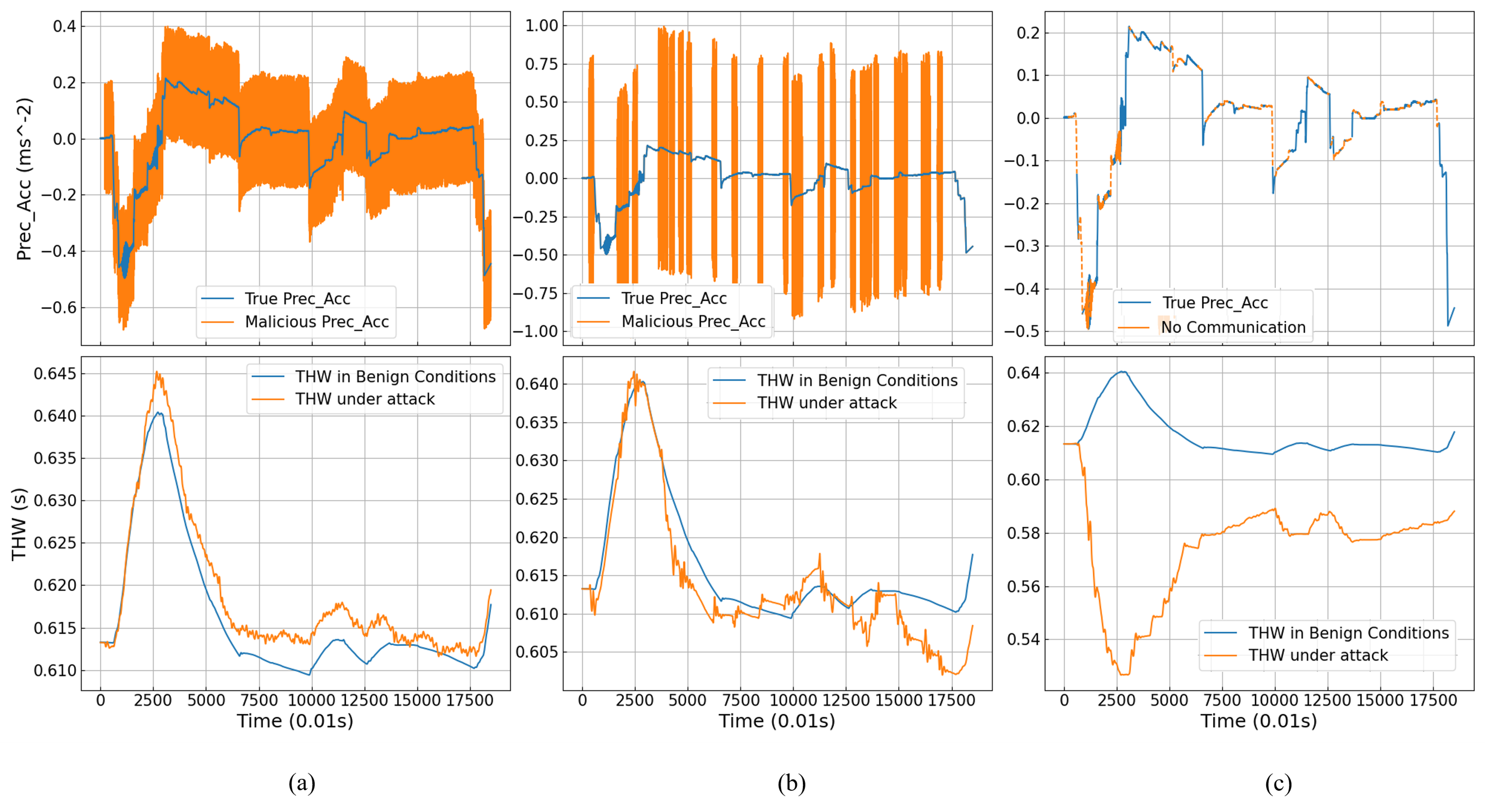}
\caption{{\small Impact of Random Bias and Delivery Prevention Attacks. (a) Random continuous bias \{-0.2, 0.2\}. (b) Random clustered bias \{-0.8, 0.8 \}. (c) Intermittent communication.}}
\label{fig:random_attacks}
\end{figure*}

Our attack orchestration framework (Section \ref{subsec:attacks}) enables systematic exploration of V2V attack space.  However, to evaluate a resiliency solution we must also comprehend the \textit{impact} of these attacks.  Note that the impact depends not only on the magnitude of the bias (deviation from normal) but also the frequency: an attack with a small bias, but performed for a long duration, can cause a significant impact on the victim vehicle. Based on the taxonomy, we perform extensive experiments across the attack space to comprehend the impact of different categories of attacks.  
%For evaluating impacts, THW ($t_{\mbox{gap}}$) is a natural metric quantifying the risk of collision or the extent of efficiency degradation; an erratic change in $t_{\mbox{gap}}$ can also indicate possibility of string instability.  

Figs.~\ref{fig:continuous_attacks}, \ref{fig:cluster_attacks}, \ref{fig:discrete_attacks}, and \ref{fig:random_attacks} show the results of impact analysis for $12$ attack instances. These are specifically chosen attack scenarios that result in a perceptible impact on the target vehicle, while remaining stealthy. The attacks either involve smaller deviations from ground truth or infrequent malicious activity, making them hard to detect. %These representative attacks have been specifically chosen since they have a perceptible impact  while requiring small deviations from ground truth or infrequent malicious activity making them hard to detect.  
%Fig.~\ref{fig:continuous_attacks} presents three representative continuous attacks involving linear mutation and sinusoidal mutation of ground truth; Fig.~\ref{fig:cluster_attacks} shows cluster attacks where malicious activity occurs for a duration $2$s once every $150$s; Fig.~\ref{fig:discrete_attacks} presents discrete attacks of varying frequencies. Fig.~\ref{fig:random_attacks} presents attacks involving random or intermittent communication.  
In addition to the impact of the individual attack instances, we can make several cumulative conclusions.  In particular, discrete attacks generally have lesser impact on the vehicle compared to cluster or continuous attacks. Furthermore, attacks that involve systematic mutation of ground truth (linear or sinusoidal) have significantly more impact on the target vehicle than attacks involving random mutation. In Section \ref{sec:ResiliencyEval}, we will demonstrate the efficacy of RACCON under a similar set of attack instances.

\section{Anomaly Detection Threshold}
\label{sec:AnomalyThresh}

RACCON resiliency depends on the choice of the anomaly threshold: a threshold higher than optimal may lead to reduced detection accuracy, while a lower threshold may lead to increased false alarms in detection.  High degree of false alarms results in inefficient invocation of RACCON's Plausibility checker. %In RACCON, false alarms invoke the plausibility checker resulting in unnecessary computation. 
Although plausibility checking computation is  lightweight, the cumulative overhead can become significant since on-road vehicles operate mostly under benign conditions.  An optimal threshold  would  enable  safety  as well as efficiency  under  adversarial  scenarios  while  incurring  minimal performance overhead in benign conditions.  Our threshold estimation methodology works in three stages: 

\begin{enumerate}
\item Identify an acceptable threshold range for adversarial scenarios.
\item Compute an approximate threshold value within the range by accounting for performance overhead under benign conditions.
\item Fine-tune the value to optimize for detection subversion attacks.  (See Section \ref{sec:DetectorSubversion}.)
\end{enumerate}

\medskip

\begin{figure*}
\centering
\subfloat[]{
  \includegraphics[width= 0.23\textwidth]{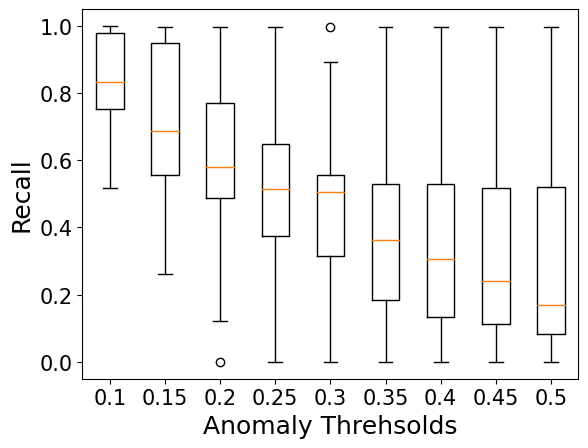} 
}
\subfloat[]{
  \includegraphics[width= 0.23\textwidth]{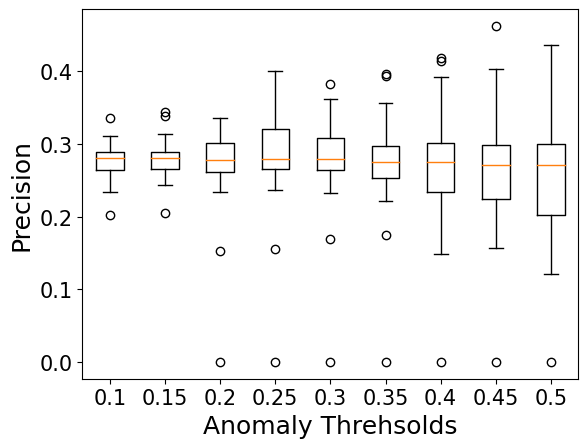}
}
\subfloat[]{
  \includegraphics[width= 0.23\textwidth]{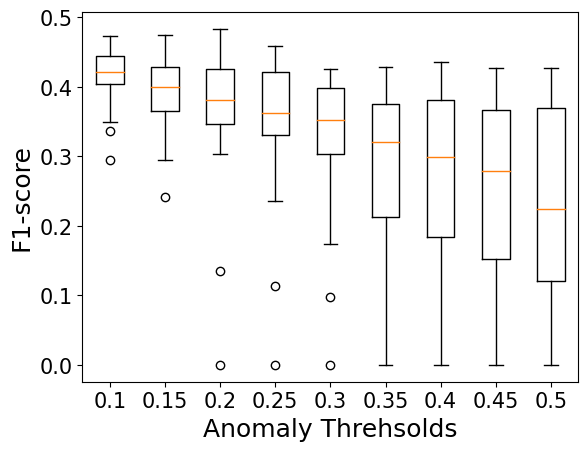} 
}
\subfloat[]{
  \includegraphics[width= 0.23\textwidth]{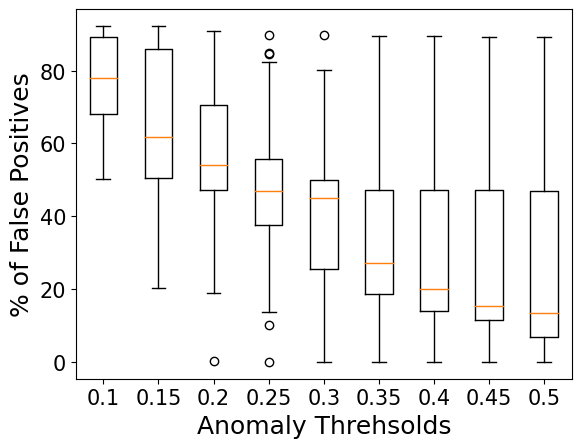} 
}
\caption{{\small Distribution Box-plots of Detection metrics vs Anomaly Threshold for $24$ Driving Environments. Plots (a) through (c) show the distribution of recall, precision and f1-score under a sinusoidal attack. Plot (d) shows the distribution of false positive percentage in benign conditions.}}
\label{fig:thresholds}
\end{figure*}

\noindent {\bf Computing Acceptable Threshold Range.}   We use three detection metrics: \textit{recall}, \textit{precision}, and \textit{f1-score}, to estimate the quality of resiliency under  attacks.  A high precision value reflects smaller percentage of false alarms while a high recall reflects  smaller percentage of undetected anomalies. A high f1-score (computed as the harmonic mean of recall and precision) indicates a combination of high precision as well as recall. We prioritize recall over precision  since it is important to capture any anomaly that can possibly cause an undesired impact.  Fig.~\ref{fig:thresholds}(a), (b), and (c) show the distribution box-plots of the three detection metrics over all $24$ environments. The evaluation is carried out under a clustered sinusoidal attack corrupting about $25\%$ of the V2V messages.  This attack is representative since it includes characteristics of both discrete and continuous attacks, and incorporates both positive and negative biases within the same attack instance.  Note that recall degrades as the anomaly threshold increases from $0.1$ to $0.5$.  The best recall values (close to $1$) are observed for thresholds in the range $0.1$-$0.2$; however, the corresponding precision values are only $0.25$-$0.35$, indicating higher number of false alarms.   Consequently, f1-scores reach an optimal value ($\sim0.4$) for smaller values of the threshold ($0.05$-$0.25$) but decrease as the threshold increases. 

\medskip

\begin{remark}
Observe from Fig.~\ref{fig:thresholds} that the f1-score boxes are not tightly packed around the mean, implying that the optimal anomaly threshold (based on f1-score) can vary across environments.  Consequently, RACCON supports on-the-fly adjustment of  threshold based on the current environment, using parameters from maps (\eg, location, terrain, etc.), ambient weather, and clocks. 
\end{remark}

% A smaller threshold would enable capturing a wide range of anomalies (favoring higher recall) but may also result in additional overhead as a result of false alarms (lower precision), and consequently higher performance overhead.  

%\subsection{Precision vs Recall Trade-off}
%As shown in Fig.~\ref{fig:thresholds}, we analyze three detection metrics:  \textit{recall}, \textit{precision}, and \textit{f1-score}, for different choices of anomaly thresholds evaluated in the $24$ environments.  A high precision value reflects smaller percentage of false alarms while a high recall reflects better coverage meaning smaller percentage of anomalies going undetected. We prioritize recall over precision  since it is important to capture any anomaly that can possibly cause an undesired impact.

%Precision in the context of RACCON anomaly detection indicates the relevance of the input samples detected as anomalies. A high precision value reflects smaller percentage of false alarms. On the other hand, recall metric for RACCON indicates the comprehensiveness of the anomaly detection. A high recall value reflects better coverage meaning smaller percentage of anomalies going undetected by RACCON. We prioritize better recall over precision in this application since it is important to capture any anomaly that can possibly cause an undesired impact on the target ego vehicle in terms of safety or efficiency.

%RACCON is evaluated under an orchestrated sinusoidal attack (see Section \ref{subsec:attacks}) corrupting about $25\%$ of the total V2V messages received. As the anomaly threshold increases from $0.1$ to $0.5$, recall degrades.  The best recall values (close to $1$) are seen for smaller threshold values of $0.1$-$0.2$.  However,  the corresponding precision values are around $0.25-0.35$, indicating higher number of false alarms. \textbf{\textit{As a result, f1-scores computed as the harmonic mean of precision and recall, reach an optimal value approximately around $0.4$ for smaller values of the threshold and decrease as the threshold increases.}} A smaller threshold would enable capturing a wide range of anomalies (favoring higher recall) but may also result in additional overhead as a result of false alarms (lower precision), and consequently higher performance overhead.  
%% Sandip:  The following two sentences need clarification.
%This overhead remains tolerable due to the light-weight on-board architecture of RACCON.  The detection and mitigation operations obey the platform requirements and consume limited computation resources.  

\medskip
 
\noindent {\bf Performance Overhead in Benign Conditions.}
%On-road vehicles primarily operate in benign conditions. Consequently, if the number of false alarms is significant then RACCON mitigation is invoked for extended duration resulting in a significant performance overhead.  Since a low anomaly threshold would result in a higher false alarm and vice versa, it is critical to study this correlation to identify the optimum threshold value.   
 Fig.~\ref{fig:thresholds}(d)  illustrates the distribution of false positives under benign conditions for thresholds ranging $0.1$-$0.5$.  Since larger thresholds result in low recall (see above), values larger than $0.5$ are disregarded.   As with f1-score, thresholds in  the range $0.05$-$0.25$ have a high variance, indicating fluctuation with changing driving environment.
The optimal anomaly threshold is selected by balancing the trade-off between better coverage under attack conditions and minimal overhead in benign conditions.

\medskip

As an example, we obtain the optimal threshold for the environment \textit{Highway-Day-Windy} as follows.  First, we determine the ballpark range  $0.1$-$0.25$ that gives the best f1-score (recall close to $1$ and precision close to $0.4$).  We eliminate  thresholds less than $0.1$  to keep the false positives  below $30$\%, refining the range to  $0.13$-$0.25$.  This is fine-tuned after evaluation under detection subversion to obtain the optimal choice  $0.15$ (Section \ref{sec:DetectorSubversion}).

\section{RACCON Resiliency Evaluation}
\label{sec:ResiliencyEval}

We performed extensive evaluation of RACCON resiliency using our flexible attack orchestration framework. Note that related work on  detecting V2V compromises (see Section \ref{sec:Related}) does not include  real-time mitigation; the only implied mitigation entails degrading to ACC (conservative controller action relying only on the trusted sensor systems). To provide a fair evaluation of RACCON, we compare it with (1) Naive CACC with no resiliency; and (2) CACC that degrades to ACC as mitigation. One way to view this evaluation is as a comparison between two extremes for safety-compromising attacks: the naive CACC controller is efficient but at the cost of safety, while degradation to ACC provides safety guarantee but at a significant efficiency cost (since ACC headway is much larger than CACC).
%While degradation to ACC may guarantee safety, it incurs a significant efficiency cost since ACC is designed to maintain much higher THW compared to CACC. 
The goal of RACCON is to enable optimal efficiency while guaranteeing safety, by maintaining THW in the range $0.55$-$0.75$s.

%In this section, we present the detailed analysis of RACCON resiliency through systematically carried out simulation experiments, accounting for attacks causing collision, efficiency degradation or string instability.

%As discussed in Section \ref{subsec:ThreatModel} illustrated in Section \ref{sec:AttackMech}, these effects are orchestrated through message mutation, fabrication or preventing delivery to the target vehicle. Additionally, we also consider the special class of detector subversion attacks. Particularly, we simulate attacks by adding a constant bias, a sinusoidal bias and a random bias, at various frequencies, to the ground truth acceleration values. We consider delivery prevention attacks by disabling communication to the target vehicle intermittently. We show representative attack scenarios of various attack categories, comparing RACCON mitigation response with: (a) A naive CACC system with no resiliency and (b) A mitigation that forces the system to simply degrade to Adaptive Cruise Control when an attack is detected.

% Describe the tables
%Table \ref{tbl:CollisionAttacks}, Table \ref{tbl:EffDegradeAttacks} and Table \ref{tbl:RandomNoComm}

\subsection{Collision and Efficiency Degradation Attacks}
\label{subsec:collision-and-degradation-attacks}

%Under a collision attack, a target ego vehicle receives malicious communication that deviates from the ground truth by a positive bias. Such communication could result in unsafe scenarios wherein the THW between the two vehicles becomes smaller than the safety threshold, potentially increasing the risk of collisions. 

\begin{figure*}
\centering
\includegraphics[width= 0.95\textwidth]{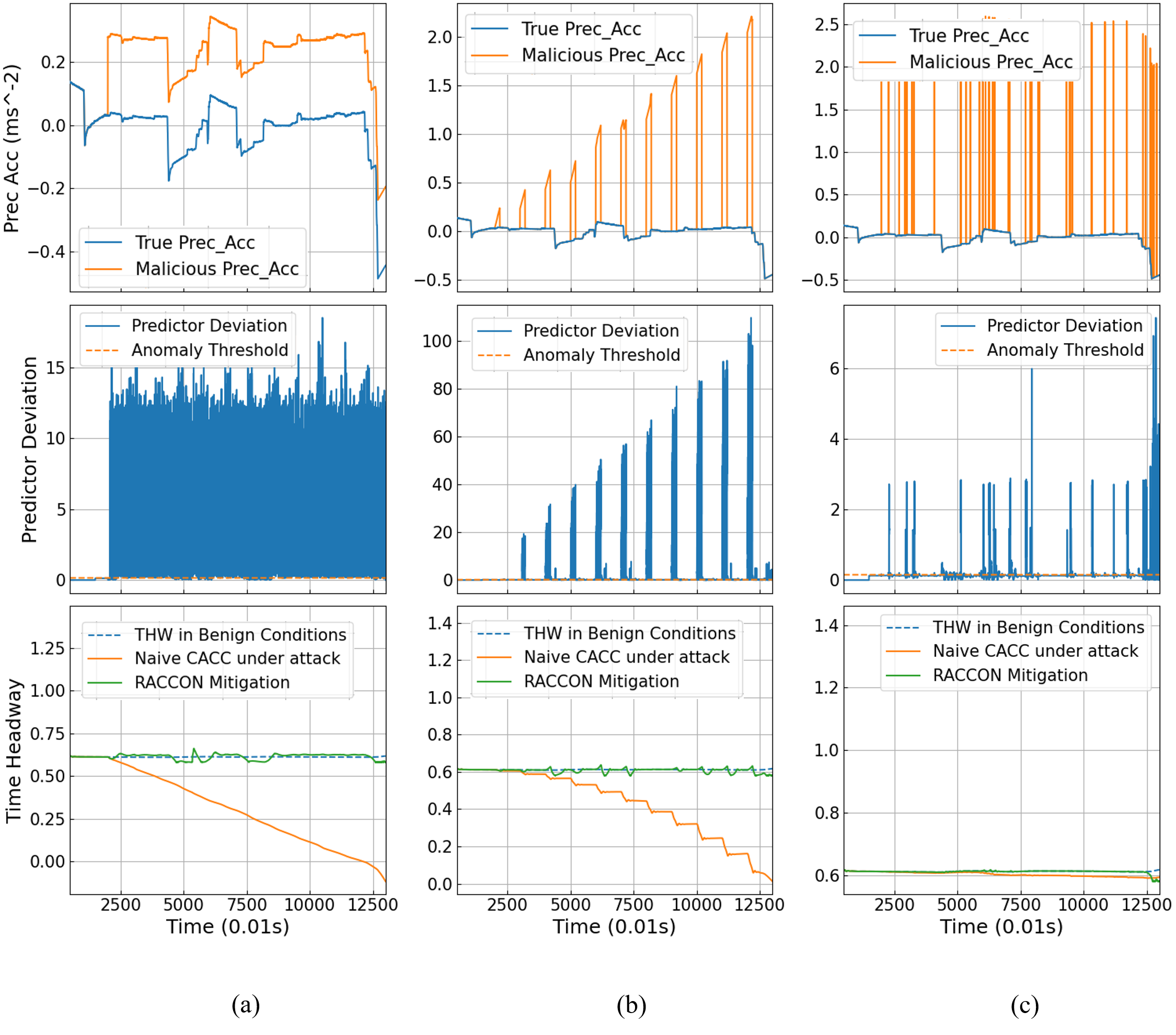}
\caption{{\small RACCON Resiliency under Sample Collision Attacks. (a) Continuous Attack constant bias +0.25. (b) Cluster Attack with linear bias +0.1t. (c) Discrete Attack with constant bias +2.5. }}
\label{fig:collision_attacks}
\end{figure*}

\begin{figure*}
\centering
\includegraphics[width= 0.95\textwidth]{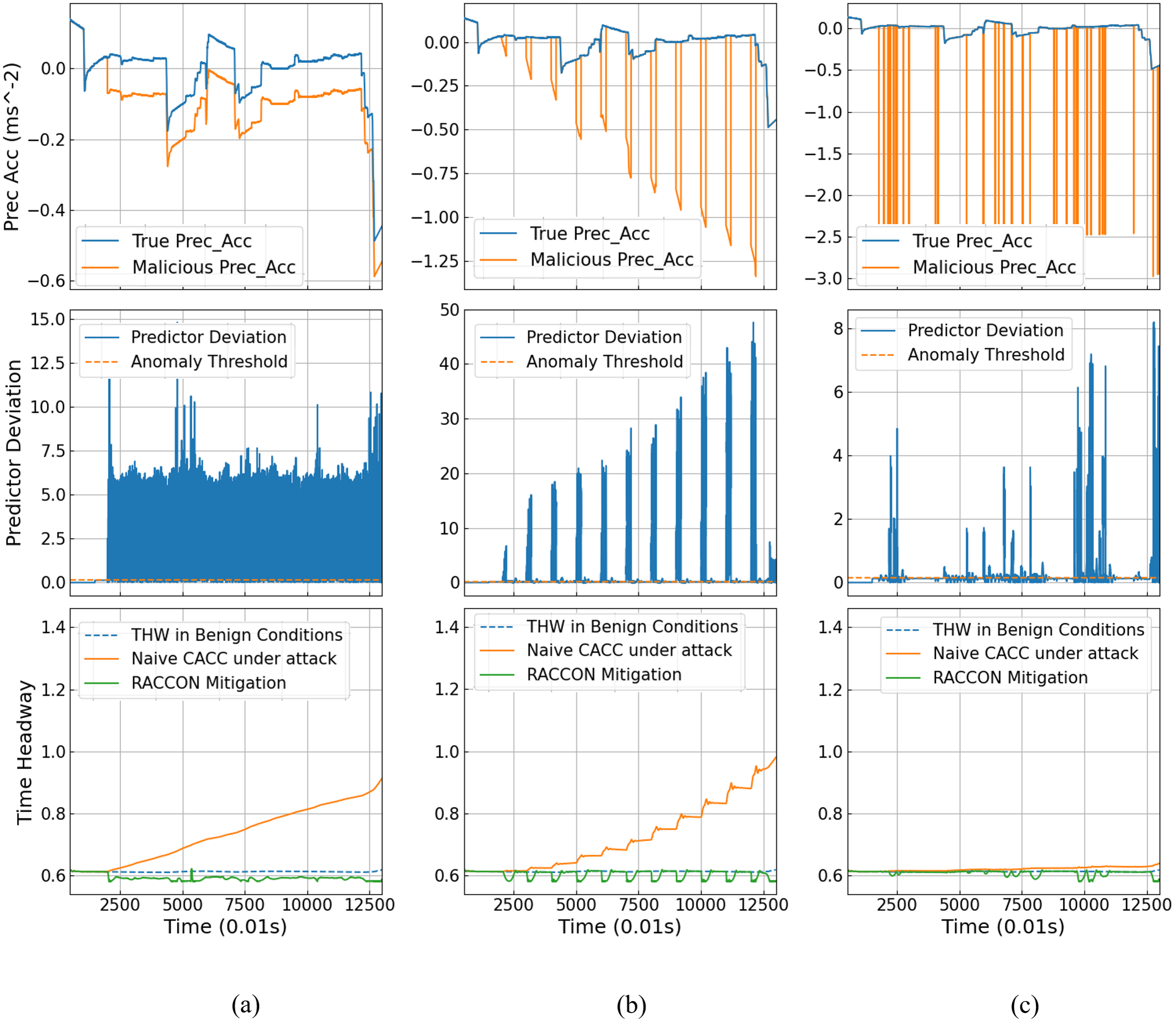}
\caption{{\small RACCON Resiliency under Sample Efficiency Degradation Attacks. (a) Continuous Attack (constant bias -0.1). (b) Cluster Attack (linear bias -0.06t). (c) Discrete Attack (constant bias -2.5).  }}
\label{fig:perf_degrad_attacks}
\end{figure*}

\begin{figure*}
\centering
\includegraphics[width= 0.95\textwidth]{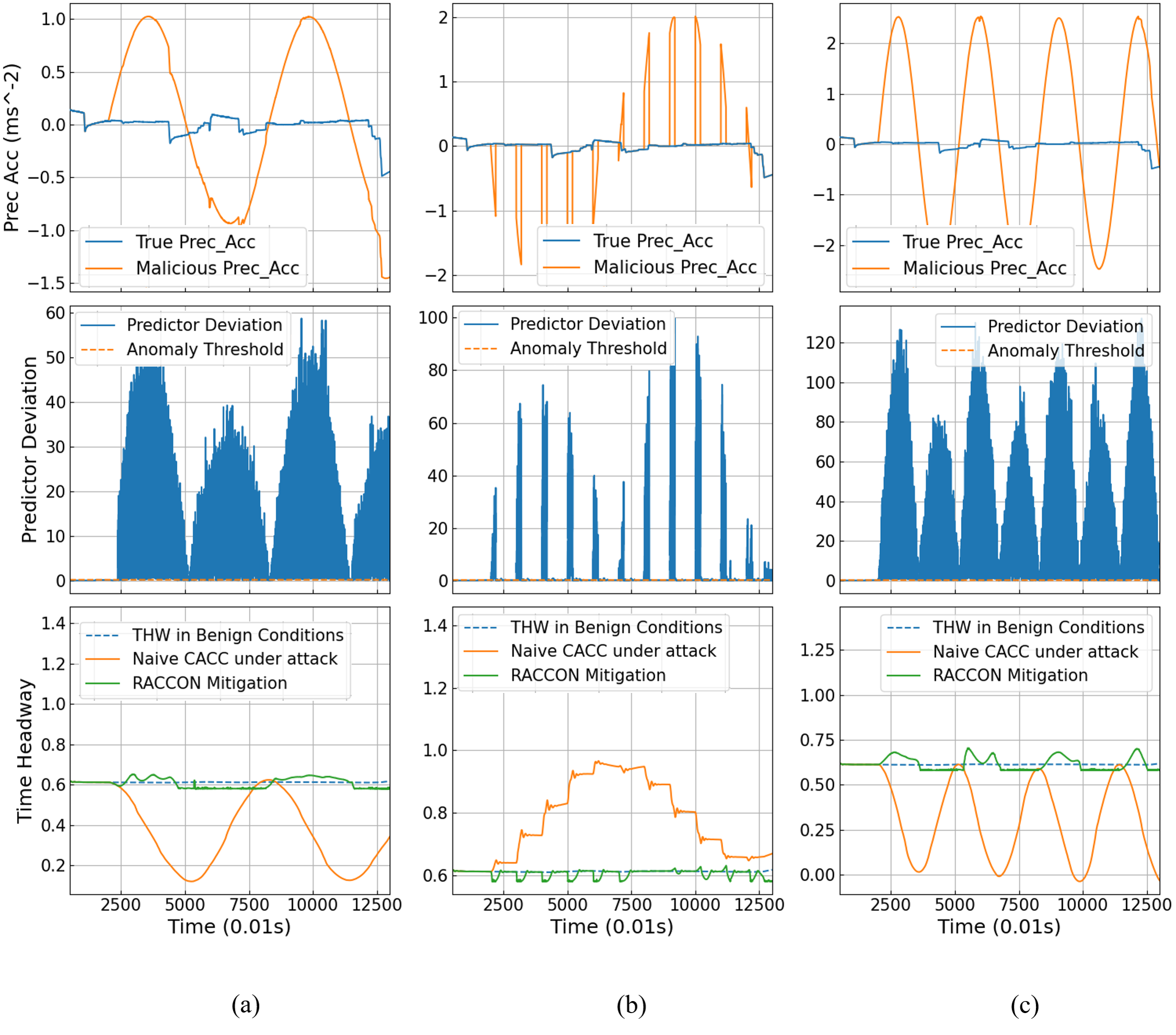}
\caption{{\small RACCON Resiliency under Sample Sinusoidal Attacks. (a) Continuous Attack (bias sin(0.1t)). (b) Cluster Attack (bias -2sin(0.3t)). (c) Continuous Attack (bias 2.5sin(0.2t)).}}
\label{fig:sinusoidal_attacks}
\end{figure*}

%A malicious V2V communication for CACC entails messages (representing preceding vehicle acceleration) that are mutated to deviate from ground trut 
%Our evaluation includes systematic orchestration of both collision attacks and efficiency degradation attacks, using biases specified as linear and sinusoidal functions of the ground truth. 
%\textcolor{red}{ Particularly, we evaluate RACCON under attacks that are hard to detect with smaller biases or infrequent malicious activity.} 
%\footnote{Sinusoidal attacks are inspired by Alotibi {\em et al.} \cite{9061150} and Jagielski {\em et al.} \cite{jagielski2018threat}.}
Tables \ref{tbl:CollisionAttacks} and \ref{tbl:EffDegradeAttacks} show the numerical results for evaluation under six representative collision and efficiency attack scenarios. Figs.~\ref{fig:collision_attacks}, \ref{fig:perf_degrad_attacks} and Fig.~\ref{fig:sinusoidal_attacks} provide visual representation of RACCON mitigation. As with Section \ref{sec:Impact}, we showcase attacks that are impactful yet hard to detect due to small biases or infrequent malicious activities. In each table, we present a comparison between RACCON, mitigation degrading to ACC, and naive CACC with no resiliency. 
Tabular entries indicate the amount of time (as percentages of total driving time) during which the vehicle experiences THW values falling within a certain range. Based on these results we make the following observations.  
\begin{itemize}
    \item {\bf Collision Attacks:} RACCON successfully mitigates the collision attacks, maintaining THW within the optimal range of $0.55$-$0.75$s at all times. CACC without any resilience results in unsafe headway of less than $0.55$s, and eventually, collision in some cases. Degrading to ACC prevents collisions, but THW is above $0.75$s for over $40\%$ of the attack duration. 
    \item  {\bf Efficiency Degradation Attacks:} With RACCON, the maximum THW is around $0.65$s. Without  resilience,  THW reaches  $1.8$s.  Degrading to ACC also results in  THW as high as $1.5$s.
\end{itemize}

%We carried out systematic collision attacks (\ie, attacks involving malicious communications that deviate from ground truth by a positive bias) through our attack orchestration paradigm.  Table \ref{tbl:CollisionAttacks} shows the results for six representative attack scenarios with varying degrees of stealth.  The malicious communication involve a linear function of the ground truth in attacks  (1), (2) and (3) and a sinusoidal function in attacks (4), (5) and (6).  Obviously, the effect of the attack is an unsafe THW for naive CACC.  Observe that under each attack scenario, RACCON successfully mitigates the attacks while maintaining the THW within $0.75$s for the entire duration of the attack.  Degrading to ACC under attack also prevents collisions, but the resultant THW is above $0.75$s for over $40$\% of the duration of the attack in each scenario indicating significant efficiency cost resulting from overly conservative mitigation.

\begin{table*}
\centering
\caption{{\small Resiliency Evaluation under Collision Attacks}}
\label{tbl:CollisionAttacks}
\resizebox{\textwidth}{!}{%
\begin{tabular}{cccccccccc}
\toprule
& \multicolumn{9}{c}{Spurious communication: Linear function of ground truth } \\
\midrule
& \multicolumn{3}{c}{ Continuous Attack (linear bias= 0.3t) } & \multicolumn{3}{c}{ Cluster Attack (constant bias= +0.8) } & \multicolumn{3}{c}{ Discrete Attack (constant bias= +2.0) } \\
\cmidrule(lr){2-4}
\cmidrule(lr){5-7}
\cmidrule(lr){8-10}
& RACCON & Degrade ACC & Naive CACC & RACCON & Degrade ACC & Naive CACC & RACCON & Degrade ACC & Naive CACC \\
\midrule
THW $<$ 0.55s & 0\% & 0\% & 84.54\% & 0\% & 0\% & 73.83\% & 0\% & 0\% & 0\% \\
THW: $\{0.55-0.75s\}$ & 100\% & 54.01\% & 15.46\% & 100\% & 51.13\% & 26.17\% & 100\% & 55.28\% & 100\% \\
THW $>$0.75s & 0\% & 45.99\% & 0\% & 0\% & 48.86\% & 0\% & 0\% & 44.72\% & 0\% \\
Collision & No & No & Yes & No & No & Yes & No & No & No \\
\midrule
& \multicolumn{9}{c}{Spurious Communication: Sinusoidal function of ground truth } \\
\midrule
& \multicolumn{3}{c}{ Continuous Attack (bias= 0.5sin(0.02t)) } & \multicolumn{3}{c}{ Cluster Attack (bias= 0.8sin(0.03t)) } & \multicolumn{3}{c}{ Cluster Attack (bias= sin(0.05t)) } \\
\cmidrule(lr){2-4}
\cmidrule(lr){5-7}
\cmidrule(lr){8-10}
& RACCON & Degrade ACC & Naive CACC & RACCON & Degrade ACC & Naive CACC & RACCON & Degrade ACC & Naive CACC \\
\midrule
THW $<$ 0.55s & 0\% & 0\% & 33.03\% & 0\% & 0\% & 12.60\% & 0\% & 0\% & 3.81\% \\
THW: $\{0.55-0.75s\}$ & 100\% & 54.64\% & 66.97\% & 100\% & 54.81\% & 87.40\% & 100\% & 53.94\% & 96.19\% \\
THW $>$0.75s & 0\% & 45.36\% & 0\% & 0\% & 45.19\% & 0\% & 0\% & 46.06\% & 0\% \\
Collision & No & No & Yes & No & No & No & No & No & No \\
\bottomrule
\end{tabular}  
}
\end{table*}

\begin{table*}
\centering
\caption{{\small Resiliency Evaluation under Efficiency Degradation Attacks}}
\label{tbl:EffDegradeAttacks}
\resizebox{\textwidth}{!}{%
\begin{tabular}{cccccccccc}
\toprule
& \multicolumn{9}{c}{ Spurious communication: Linear function of ground truth } \\
\midrule
& \multicolumn{3}{c}{ Continuous (linear bias= -0.3t) } & \multicolumn{3}{c}{ Cluster (constant bias= -0.8) } & \multicolumn{3}{c}{ Discrete (constant bias= -2.0) } \\
\cmidrule(lr){2-4}
\cmidrule(lr){5-7}
\cmidrule(lr){8-10}
& RACCON & Degrade ACC & Naive CACC & RACCON & Degrade ACC & Naive CACC & RACCON & Degrade ACC & Naive CACC \\
\midrule
THW $<$ 0.55s & 0\% & 0\% & 0\% & 0\% & 0\% & 0\% & 0\% & 0\% & 0\% \\
THW: $\{0.55-0.75s\}$ & 100\% & 55.42\% & 21.55\% & 100\% & 55.25\% & 18.85\% & 100\% & 54.83\% & 100\% \\
THW $>$0.75s & 0\% & 44.58\% & 78.45\% & 0\% & 44.75\% & 81.15\% & 0\% & 45.17\% & 0\% \\
Maximum THW & 0.65s & 1.56s & 1.79s & 0.65s & 1.55s & 1.54s & 0.65s & 1.54s & 0.70s \\
\midrule
& \multicolumn{9}{c}{ Spurious Communication: Sinusoidal function of ground truth } \\
\midrule
& \multicolumn{3}{c}{ Continuous Attack (bias= -0.5sin(0.02t)) } & \multicolumn{3}{c}{ Cluster Attack (bias= -0.8sin(0.03t)) } & \multicolumn{3}{c}{ Cluster Attack (bias= -sin(0.05t)) } \\
\cmidrule(lr){2-4}
\cmidrule(lr){5-7}
\cmidrule(lr){8-10}
& RACCON & Degrade ACC & Naive CACC & RACCON & Degrade ACC & Naive CACC & RACCON & Degrade ACC & Naive CACC \\
\midrule
THW $<$ 0.55s & 0\% & 0\% & 0\% & 0\% & 0\% & 0\% & 0\% & 0\% & 0\% \\
THW: $\{0.55-0.75s\}$ & 100\% & 54.67\% & 79.97\% & 100\% & 54.14\% & 94.01\% & 100\% & 54.49\% & 98.35\% \\
THW $>$0.75s & 0\% & 45.33\% & 20.03\% & 0\% & 45.86\% & 5.99\% & 0\% & 45.51\% & 1.65\% \\
Maximum THW & 0.65s & 1.56s & 0.83s & 0.65s & 1.55s & 0.79s & 0.65s & 1.54s & 0.75s \\
\bottomrule
\end{tabular}  
}
\end{table*}

%Figure \ref{fig:collision_attacks} and Figure \ref{fig:perf_degrad_attacks} show some representative collision attacks and efficiency degradation attacks involving linear bias between malicious communication and ground truth. Figure \ref{fig:sinusoidal_attacks} shows a few sinusoidal attacks causing collisions or efficiency degradation.

\subsection{Random Communication and Delivery Prevention attacks}
\label{subsec:random-attacks}

\begin{figure*}
\centering
\includegraphics[width= 0.95\textwidth]{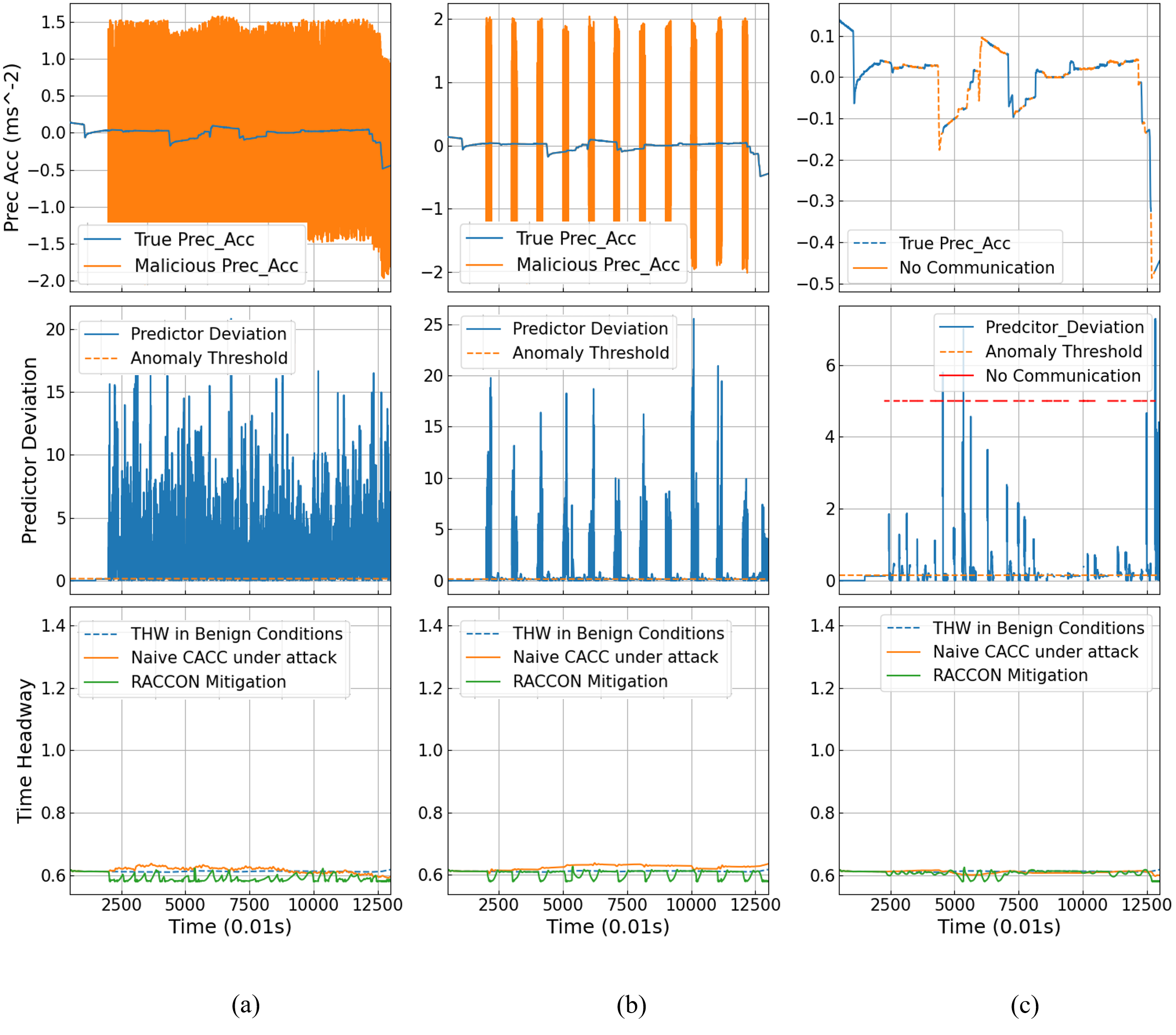}
\caption{{\small RACCON Resiliency under Random Mutation and Delivery Prevention Attacks: Comparison between RACCON and naive CACC with no resiliency, in terms of resultant THW; (a) Continuous Attack (random bias {-1.5, 1.5}); (b) Cluster Attack (random bias {-2.0, 2.0}); (c) Intermittent communication.}}
\label{fig:random_resiliency}
\end{figure*}

We also studied effects of random message mutation and delivery prevention (Table \ref{tbl:RandomNoComm} and Fig.~\ref{fig:random_resiliency}). 
The results show the importance of the thorough attack impact analysis we carried out.
Recall from Section \ref{sec:Impact} that these attacks have much less impact than Collision and Efficiency Degradation attacks.  A critical aspect of resiliency evaluation is to ensure it does not incur high mitigation overhead.  
%Table \ref{tbl:AttackMech} presents the results under six representative attacks involving random and intermittent communication, and Fig.~\ref{fig:random_resiliency} provides a pictorial representation of a few instances.  We conclude that such attacks have significantly less impact on the application than attacks involving targeted bias. 
Both RACCON and naive CACC maintain  $t_{\mbox{gap}}$ within the ideal range at all times; however, degrading to ACC incurs significant efficiency loss.

%\textcolor{red}{Although these attacks result in very little perceptible impact on the naive CACC system, mitigation degrading to ACC still incurs a significant efficiency loss. RACCON on the other hand maintains the THW in the optimal range at all times. By design, RACCON applies corrections to the system in proportion to the projected impact of an anomaly. This unique feature of RACCON minimizes efficiency loss under mitigation of any attack. Furthermore, the overhead due to mitigation under false positives is minimal as a result.}
%We studied the effect of malicious communication that deviates in a random fashion with respect to the ground truth. We also orchestrated attacks that prevent communication delivery to the target vehicle. 
%Table \ref{tbl:RandomNoComm} presents the observations under six different attacks that involve random or intermittent communication. We conclude that such attacks have considerably less impact on the target system than attacks that involve systematic mutation of the ground truth (for instance, the attacks considered in the above subsections). Figure \ref{fig:random_resiliency} shows a few representative examples of such attacks. With RACCON, the resultant $t_{\mbox{gap}}$ remains within the ideal range at all times.

\begin{table*}
\centering
\caption{{\small Resiliency Evaluation under Random Mutation and Delivery Prevention Attacks}}
\label{tbl:RandomNoComm}
\resizebox{\textwidth}{!}{%
\begin{tabular}{cccccccccc}
\toprule
& \multicolumn{9}{c}{ Random Mutation Attacks } \\
\midrule
& \multicolumn{3}{c}{ Continuous (random bias={-2.0,2.0}) } & \multicolumn{3}{c}{ Cluster (random bias={-2.0,2.0})} & \multicolumn{3}{c}{ Discrete (random bias={-2.0,2.0}) } \\
\cmidrule(lr){2-4}
\cmidrule(lr){5-7}
\cmidrule(lr){8-10}
& RACCON & Degrade ACC & Naive CACC & RACCON & Degrade ACC & Naive CACC & RACCON & Degrade ACC & Naive CACC \\
\midrule
THW $<$ 0.55s & 0\% & 0\% & 0\% & 0\% & 0\% & 0\% & 0\% & 0\% & 0\% \\
THW: $\{0.55-0.75s\}$ & 100\% & 54.07\% & 100\% & 100\% & 55.69\% & 100\% & 100\% & 55.20\% & 100\%  \\
THW $>$0.75s & 0\% & 45.93\% & 0\% & 0\% & 44.31\% & 0\% & 0\% & 44.80\% & 0\% \\
Max THW &0.65 & 1.54 & 0.73 & 0.65 & 1.55 & 0.65 &0.65 & 1.54 &0.65 \\
\midrule
& \multicolumn{9}{c}{Delivery Prevention Attacks} \\
\midrule
& \multicolumn{3}{c}{ Intermittent (frequency= 0.2Hz, duration=1.5s) } & \multicolumn{3}{c}{ Intermittent (frequency= 0.1Hz, duration=2s)} & \multicolumn{3}{c}{ Intermittent (frequency= 0.2Hz, duration=5s) } \\
\cmidrule(lr){2-4}
\cmidrule(lr){5-7}
\cmidrule(lr){8-10}
& RACCON & Degrade ACC & Naive CACC & RACCON & Degrade ACC & Naive CACC & RACCON & Degrade ACC & Naive CACC \\
\midrule
THW $<$ 0.55s & 0\% & 0\% & 0\% & 0\% & 0\% & 0\% & 0\% & 0\% & 3.29\% \\
THW: $\{0.55-0.75s\}$ & 100\% & 54.86\% & 100\% & 100\% & 54.88\% & 100\% & 100\% & 54.92\% & 96.71\% \\
THW $>$0.75s &0\%  &45.14\% & 0\% & 0\% & 45.12\% & 0\% & 0\% & 45.08\% & 0\%\\
Max THW & 0.65 & 1.54 & 0.65 & 0.65 & 1.54 &0.65 & 0.65 & 1.54 & 0.66 \\
\bottomrule
\end{tabular}  
}
\end{table*}

\subsection{N-Day Attacks}
Attacks orchestrated in Sections \ref{subsec:collision-and-degradation-attacks} and \ref{subsec:random-attacks} systematically cover the taxonomy discussed in Section \ref{subsec:Taxonomy}. Since our taxonomy comprehensively represents the whole V2V attack spectrum, it is established from our evaluation results that RACCON is robust against any arbitrary V2V attack under the threat model, including both known ($N$-day) and unknown ($0$-day) attacks.  Nevertheless, it is illustrative to directly evaluate RACCON against  some known attacks.  In this section, we consider three well-known attacks, \eg, Man-in-the-Middle (MITM), Denial-of-Service (DoS) through Jamming, and DoS through Flooding.  

\begin{itemize}
\item {\bf MITM Attack:}  We instantiate an MITM adversary that mutates the preceding vehicle acceleration values by adding a continuous sinusoidal bias, using the function $0.8\sin 0.05t$.
\item {\bf DoS through Jamming:} We implement a DoS attack in which the adversary jams the communication channel, preventing delivery of (legitimate) V2V messages.  The channel is jammed for $2$ seconds once every $20$ seconds.
\item {\bf DoS through Flooding:}. The adversary floods the communication channel with fabricated packets that interfere with delivery of legitimate communication.  We add fabricated packets in bursts, once every $10$ seconds, for a duration of $2$ seconds. 
\end{itemize}

\begin{figure*}
\centering
\includegraphics[width= 0.95\textwidth]{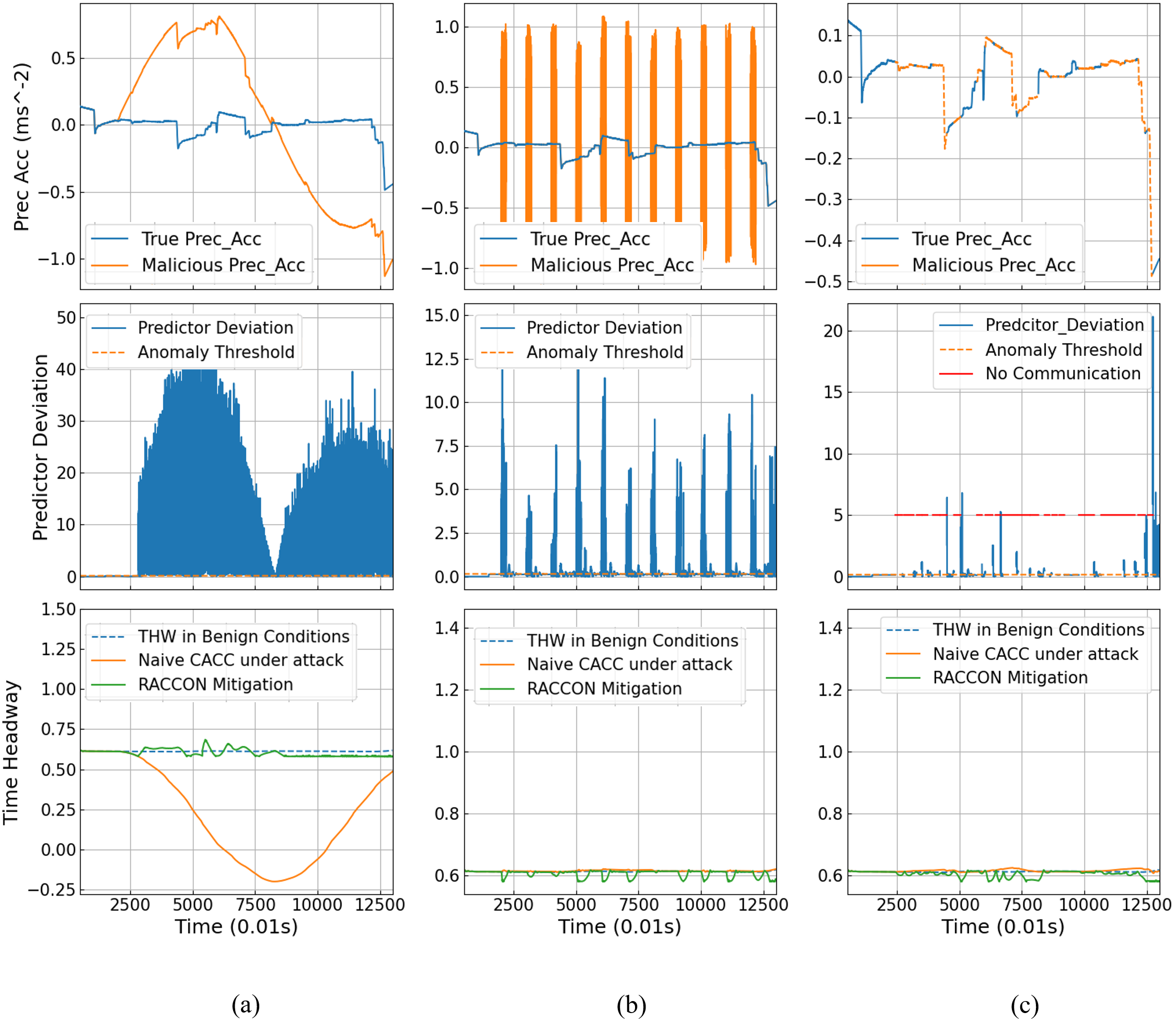}
\caption{{\small RACCON Resiliency under Representative N-day Attacks. (a) MITM Attack with continuous sinusoidal bias. (b) Flooding Attack with cluster random bias. (c) DoS Attack with intermittent communication. }}
\label{fig:n-day_attacks}
\end{figure*}

Fig.~\ref{fig:n-day_attacks} illustrates RACCON mitigation efficacy under these attacks. It maintains $t_{\mbox{gap}}$ close to ideal at all times, while CACC without resiliency results in $t_{\mbox{gap}}$ of less than $0.55$s for MITM. Mitigation based on fallback to ACC results in significant efficiency degradation for the jamming attack.

\section{Detector Subversion}
\label{sec:DetectorSubversion}

The fact that RACCON is an ML-based framework can make it vulnerable to adversaries subverting the learning and prediction systems themselves.  Such adversaries can create anomalous data that is nevertheless accepted as normal by the detector, thereby bypassing any mitigation against the attack.  We call these attacks {\em detector subversion}.  
%We made extensive evaluation of the impact of these attacks on the safety and efficiency of RACCON.

%\begin{remark}
%One can argue that detector subversion which is not included in the taxonomy suggests a hole in taxonomy itself.  However, the taxonomy is only comprehensive for attacks that {\em impact} the ego vehicle.  Deviations resulting from a successful detector subversion attack must be very small (less than threshold) to be classified as normal; consequently, although these attacks can ``fool'' the Predictor, they would not result in a perceptible effect on the safety and efficiency of the application for appropriate threshold choices.
%\end{remark}

%The key intuition behind RACCON resiliency to detector subversion is that for a (sequence of) malicious communications to be misclassified as normal by RACCON, the deviation produced in the response of the ego vehicle as a result of these communications must be less than the anomaly threshold.  On the other hand, a very small deviation from threshold would not result in any perceptible impact on safety or efficiency of the application.  
Obviously, a very low selection of anomaly threshold can ensure high robustness against detector subversion. However, recall from Section \ref{sec:AnomalyThresh} that a low anomaly threshold can result in high false alarms.  Consequently, we fine-tune the threshold value within the ballpark range obtained from Section \ref{sec:AnomalyThresh}, balancing the trade-off. We use the following parameters in our analysis.   

%Since RACCON is a data-driven framework based on anomaly detection through machine learning, one potential risk is that an adversary can attempt to subvert the detection system. The adversary can send anomalous data that nevertheless is accepted as ``normal'' by the detector, thereby bypassing any mitigation against the attack.  We call these attacks {\em detector subversion attacks}.  We made extensive evaluation of the impact of these attacks on the safety and efficiency of the RACCON system. 

%Robustness of RACCON against subversion attacks relies on the optimal choice of the anomaly threshold. The smaller the anomaly threshold, more robust is the security system against subversion attacks, since it enables capturing minute variations from normal behavior. However, the amount of false positives, {\ie normal communication tagged as anomalous}, increases as the anomaly threshold becomes smaller. A high degree of false positives leads to increased computation overhead, due to the invocation of the mitigation system even under benign operating conditions. Therefore, anomaly threshold should be carefully determined to strike the right balance between the desired robustness against subversion attacks and the resultant computation overhead. 

%We define the following parameters to evaluate the choice of anomaly threshold for RACCON. The goal is to determine the optimal anomaly threshold which enables the detection of every attack beyond the tolerable bias, while keeping the the number of  false positives small.     
\begin{itemize}
    \item \textit{Tolerable Bias:} This is the maximum bias added to the ground truth, beyond which there is a perceptible impact on the target vehicle's safety or efficiency.
    \item \textit{Subversion Detectability Index:} This is the  minimum bias added to ground truth, that can be successfully captured by the detection system.
    \item \textit{False Positives in Benign Conditions:} This is the  percentage of normal communication messages, incorrectly tagged as anomalies by RACCON in benign operating conditions.  
\end{itemize}
The goal is to determine the optimal anomaly threshold which enables the detection of every attack beyond the tolerable bias, while keeping the the number of  false positives small.
%We experimented with different values of anomaly detection thresholds and recorded the observations of the above mentioned metrics. The goal is to determine the optimal anomaly threshold which enables the detection of every attack beyond the tolerable bias, while keeping the false positive \% small.     

Table \ref{tbl:DetSubversion} presents results for threshold choices for a representative driving environment, \textit{Highway-Day-Windy}.  Recall from Section \ref{sec:AnomalyThresh} that we determined the approximate optimal threshold range for this environment to be $0.12$-$0.25$.  To fine-tune for resiliency under detector subversion, we determine the tolerable bias for attacks of varying stealth factor; note that it is much smaller for a continuous attack ($0.04$) than a discrete attack ($5.0$).  For optimal threshold, the subversion detectability index  should be less than the tolerable bias for each class of attack.  The highlighted row shows the optimal choice of the anomaly threshold ($0.15$), since it has the minimum fraction of false positives out of all the choices providing acceptable subversion detectability.

\begin{table*}
\centering
\caption{{\small Anomaly Threshold and Subversion Detectability under Attacks of Varying Stealth Factor.}}
\label{tbl:DetSubversion}
\resizebox{\textwidth}{!}{%
\begin{tabular}{cccccccc}
\toprule
\multirow{3}{*}{\shortstack{ \\ Anomaly \\ Threshold }} & \multirow{3}{*}{\shortstack{ \\ False Positives \\ Benign Condition }} & \multicolumn{6}{c}{Subversion Detectability Index} \\
\cmidrule{3-8}
& & \multicolumn{2}{c}{ Continuous (Tolerable bias: 0.04) } & \multicolumn{2}{c}{ Cluster (Tolerable bias: 0.1) } & \multicolumn{2}{c}{ Discrete (Tolerable bias: 5.0) } \\
\cmidrule(lr){3-4}
\cmidrule(lr){5-6}
\cmidrule(lr){7-8}
& & Min. constant bias & Min. sinusoidal bias & Min.  constant bias & Min. sinusoidal bias & Min. constant bias & Min. sinusoidal bias \\
\midrule
0.25 & 0\% & 0.35 & 0.25sinft & 0.4 & 0.35sinft & 0.5 & 3sinft \\
0.2 & 2.96\% & 0.3 & 0.2sinft & 0.3 & 0.3sinft & 0.35 & 1sinft \\
0.18 & 10.74\% & 0.3 & 0.2sinft & 0.3 & 0.3sinft & 0.35 & 0.35sinft \\
\rowcolor{maroon!10} 0.15 & 11.91\% & 0.01 & 0.01sinft & 0.03 & 0.02sinft & 0.25 & 0.25sinft \\
0.13 & 21.2\% & 0 & 0 & 0.0001 & 0.0001sinft & 0.01 & 0.01sinft \\
0.12 & 58.1\% & 0 & 0 & 0 & 0 & 0 & 0 \\
\bottomrule
\end{tabular} 
}
\end{table*}

\section{Related Work and Discussion}
\label{sec:Related}

Automotive security research has been traditionally focused on in-vehicle vulnerabilities or adversaries exploiting the lack of secure communication \cite{checkoway2011,koscher2010,miller2015}. Machine learning has primarily been used for computer vision modules to improve on-board perception \cite{Uricar:2019,Tian:2018:icse} or for securing in-vehicle networks, \eg, CAN bus \cite{CAN_Anomaly_LSTM,CAN_Anomaly_2}.  With the emergence of CAV systems, recent research has focused on security of cooperative and safety applications such as platooning \cite{platoon-security}, intersection management \cite{buzachis2018secure}, collision avoidance,  emergency vehicle warning, lane merge and turn conflict warning, etc. \cite{blum2008denial},\cite{kim2014taxonomy}

Since CACC serves as a foundation of a variety of CAV applications, significant attention has been given towards detection of attacks on CACC.  This research primarily involves application of control theory or machine learning solutions.  Abdo {\em et al.} \cite{abdo2019application} present a survey on application level communication attacks on CACC and their adverse impacts on the target vehicles.   Liu {\em et al.} \cite{8515151}, Parkinson {\em et al.} \cite{7872388} and AbdAllah {\em et al.} \cite{10.1145/3123779.3123794} discuss the challenges in CACC security and provide research directions.  Biron {\em et al.} \cite{CACCFaultDetection} and  Dutta {\em et al.} \cite{SecForSafety} use approaches based on control theory to detect and correct adversarial sensor-based attacks on CACC. Heijden {\em et al.} \cite{7881000} propose a misbehavior detection mechanism based on subjective logic, to validate the position information exchanged between vehicles.  Nunen {\em et al.} \cite{ModelPredictiveCACC} propose a control-theoretic model-predictive approach to deal with short communication failures and packet dropouts in CACC. Among machine learning approaches, Alotibi {\em et al.} \cite{9061150} propose a real-time detection mechanism for platoons, in the context of a compromised leader reporting falsified acceleration values to the following vehicles.  Iorio {\em et al.} \cite{9062798} propose a misbehavior detection approach for injection attacks on CACC, based on correlation between various vehicular motion parameters. Jagielski {\em et al.} \cite{jagielski2018threat} discuss  detection of attacks that compromise communication or manipulate the on-board sensor readings, through physics-based constraints and machine learning.   Levi {\em et al.} \cite{DBLP:journals/corr/abs-1711-01939} present an event-based anomaly detection technique for connected vehicles using Hidden Markov Models. Tiwari {\em et al.}~\cite{Tiwari:2014} describe attack features that are undetectable at individual time instances but can be detected from sequential data. 
%There has also been related work on adversarial attacks on ML-based intrusion detection systems \cite{Madry:2018,zhang:2019:iclr}. 

In spite of this extensive research,  we are not aware of any previous solution addressing detection of the spectrum of attacks explored for RACCON. Control-theoretic approaches require a detailed functional model of the adversarial action. Each attack type (\eg, flooding, jamming, etc.) requires a different detailed adversary model. In contrast, RACCON is an ML-based anomaly detection approach that only depends upon benign V2V communication data. RACCON's attack-agnostic defense is effective against the entire spectrum of V2V adversaries. On the other hand, related ML-based approaches have only been evaluated under a specific subset of attacks, \eg, linear or sinusoidal mutation attacks on acceleration values \cite{9061150,jagielski2018threat}.  
%No other related work systematically orchestrates attacks covering the entire spectrum defined by the threat model or provides a comprehensive evaluation of the detection mechanism across the attack spectrum.

%In spite of this extensive research, we are not aware of any  solution that can defend against the entire spectrum of communication attacks on CACC. Control-theoretic approaches require detailed functional modelling of adversary operation for each attack type (\eg, DoS, flooding, jamming, etc.)  In contrast, RACCON is attack agnostic in its defense; it relies purely on V2V communication data  in benign operating conditions and does not require any prior knowledge of the attack mechanisms during its training.  On the other hand, none of the related works shows the systematic orchestration of the entire attack spectrum defined by the threat model. And consequently, they do not present comprehensive evaluation of the proposed detection techniques. RACCON is extensively evaluated under a comprehensive attack taxonomy that accounts for any compromise that can be orchestrated on V2V in CACC. 
% No other approach systematically orchestrates attacks covering the entire spectrum defined by the threat model or provides a comprehensive evaluation of the detection mechanism across the attack spectrum. 

A unique aspect of RACCON is {\em real-time resiliency}, providing optimal efficiency while guaranteeing safety under adversarial conditions.  This vision has guided several components of RACCON's design and evaluation. For instance, while all related ML-based anomaly detection approaches focus on identifying discrepancies in controller \textit{inputs}, RACCON is designed to monitor the controller's \textit{response}.
%For instance, while all related ML-based anomaly detection approaches for CACC evaluate the \textit{inputs} to the controller as normal or anomalous, RACCON evaluates the \textit{response} of the ego vehicle; 
This permits RACCON to correct the erroneous response appropriately and minimize the impact of anomalous (and potentially malicious) inputs on the ego vehicle.  The need for resiliency also requires us to determine the severity and impact of the attack itself, \ie, an attack is impactful and needs mitigation if it results in the ego vehicle performing an unsafe or inefficient action.  This requirement has also led to the understanding of the trade-offs between stealth and impact, \eg, clustered and continuous attacks are more impactful than discrete attacks, and are correspondingly less stealthy. The need for real-time responses has motivated our design goals for viable ML models that satisfy automotive resource constraints. Finally, the complex trade-off between robustness and performance has guided our methodology for optimal anomaly threshold computation.

\section{Conclusion and Future Work}
\label{sec:Concl}

We have presented what we believe is the first comprehensive resiliency framework for CACC against V2V attacks.  Our work uses machine learning to predict the ego vehicle's responses, and capture communication anomalies in real-time, based on deviation between the predicted and actual responses. We also developed a robust real-time mitigation technique that can effectively nullify the adverse effects of anomalous communication.  A unique feature of this mitigation is to guarantee safety while preserving efficiency. Unlike systems that degrade to ACC in response to an anomaly, our solution enables the target vehicle to safely engage in CACC even under attack.   We have also developed one of the most comprehensive experimental frameworks for resiliency evaluation, based on a taxonomy of adversaries capturing the entirety of the V2V attack spectrum.  Our experiments clearly demonstrate the viability of RACCON as a means for providing resiliency in CACC under V2V attacks.

%We have presented an architecture for Cooperative Adaptive Cruise Control that is resilient against  attacks on V2V communications.  Our work uses machine learning to compute predicted response of the ego vehicle, and uses this prediction to determine communication anomalies and perform mitigation.  A unique feature is the ability to mitigate attacks in real-time, which enables us to use CACC even under anomaly, achieving better efficiency than systems that degrade to ACC.  We have performed extensive experimental evaluation to demonstrate the effectiveness of our approach. 

In our future work, we will explore extension of this resiliency architecture to other connected car applications.  We will also augment RACCON with existing techniques for additionally detecting sensor attacks, resulting in more robust CACC.

\bibliographystyle{abbrv}
\bibliography{security}

\end{document}